\documentclass[aps,prb,reprint,superscriptaddress]{revtex4-2}
\usepackage[T1]{fontenc}
\usepackage{times}
\usepackage{graphicx,color}
\usepackage{amsfonts,amsmath,amssymb,amsbsy}
\usepackage{float}
\usepackage[colorlinks=true, linkcolor=blue, citecolor=blue, urlcolor=blue]{hyperref}
\def\B{{\mathcal{B}}}
\def\D{{\mathcal{D}}}

\def\I{{\mathcal{I}}}
\let\mathbf=\boldsymbol

\def\blue#1{\textcolor{blue}{#1}}
\def\emph#1{\textcolor{magenta}{#1}}
\begin{document}

\title{Laminar and transiently disordered dynamics of a magnetic skyrmion pipe flow}

\author{Xichao Zhang}
\thanks{These authors contributed equally to this work.}
\affiliation{Department of Applied Physics, Waseda University, Okubo, Shinjuku-ku, Tokyo 169-8555, Japan}

\author{Jing Xia}
\thanks{These authors contributed equally to this work.}
\affiliation{Department of Electrical and Computer Engineering, Shinshu University, 4-17-1 Wakasato, Nagano 380-8553, Japan}

\author{Oleg A. Tretiakov}
\affiliation{School of Physics, The University of New South Wales, Sydney 2052, Australia}

\author{Motohiko Ezawa}
\affiliation{Department of Applied Physics, The University of Tokyo, 7-3-1 Hongo, Tokyo 113-8656, Japan}

\author{Guoping Zhao}
\affiliation{College of Physics and Electronic Engineering, Sichuan Normal University, Chengdu 610068, China}

\author{Yan Zhou}
\affiliation{School of Science and Engineering, The Chinese University of Hong Kong, Shenzhen, Guangdong 518172, China}

\author{\\ Xiaoxi Liu}
\email[Email:~]{liu@cs.shinshu-u.ac.jp}
\affiliation{Department of Electrical and Computer Engineering, Shinshu University, 4-17-1 Wakasato, Nagano 380-8553, Japan}

\author{Masahito Mochizuki}
\email[Email:~]{masa_mochizuki@waseda.jp}
\affiliation{Department of Applied Physics, Waseda University, Okubo, Shinjuku-ku, Tokyo 169-8555, Japan}

\begin{abstract}
The world is full of fluids that flow. The fluid nature of flowing skyrmionic quasiparticles is of fundamental physical interest and plays an essential role in the transport of many skyrmions. Here, we report the laminar and transiently disordered dynamic behaviors of many magnetic skyrmions flowing in a pipe channel. The skyrmion flow driven by a uniform current may show a lattice structural transition. The skyrmion flow driven by a non-uniform current shows a dynamically varying lattice structure. A large uniform current could result in the compression of skyrmions toward the channel edge, leading to the transition of the skyrmion pipe flow into an open-channel flow with a free surface. Namely, the width of the skyrmion flow could be adjusted by the driving current. Skyrmions on the free surface may form a single shear layer adjacent to the main skyrmion flow. In addition, although we focus on the skyrmion flow dynamics in a clean pipe channel without any pinning or defect effect, we also show that a variation of magnetic anisotropy in the pipe channel could lead to more complicated skyrmion flow dynamics and pathlines. Our results reveal the fluid nature of skyrmionic quasiparticles that may motivate future research on the complex flow physics of magnetic textures.
\end{abstract}

\date{October 3, 2023}


\maketitle

\section{Introduction}
\label{se:Introduction}

\begin{figure*}[t]
\centerline{\includegraphics[width=0.95\textwidth]{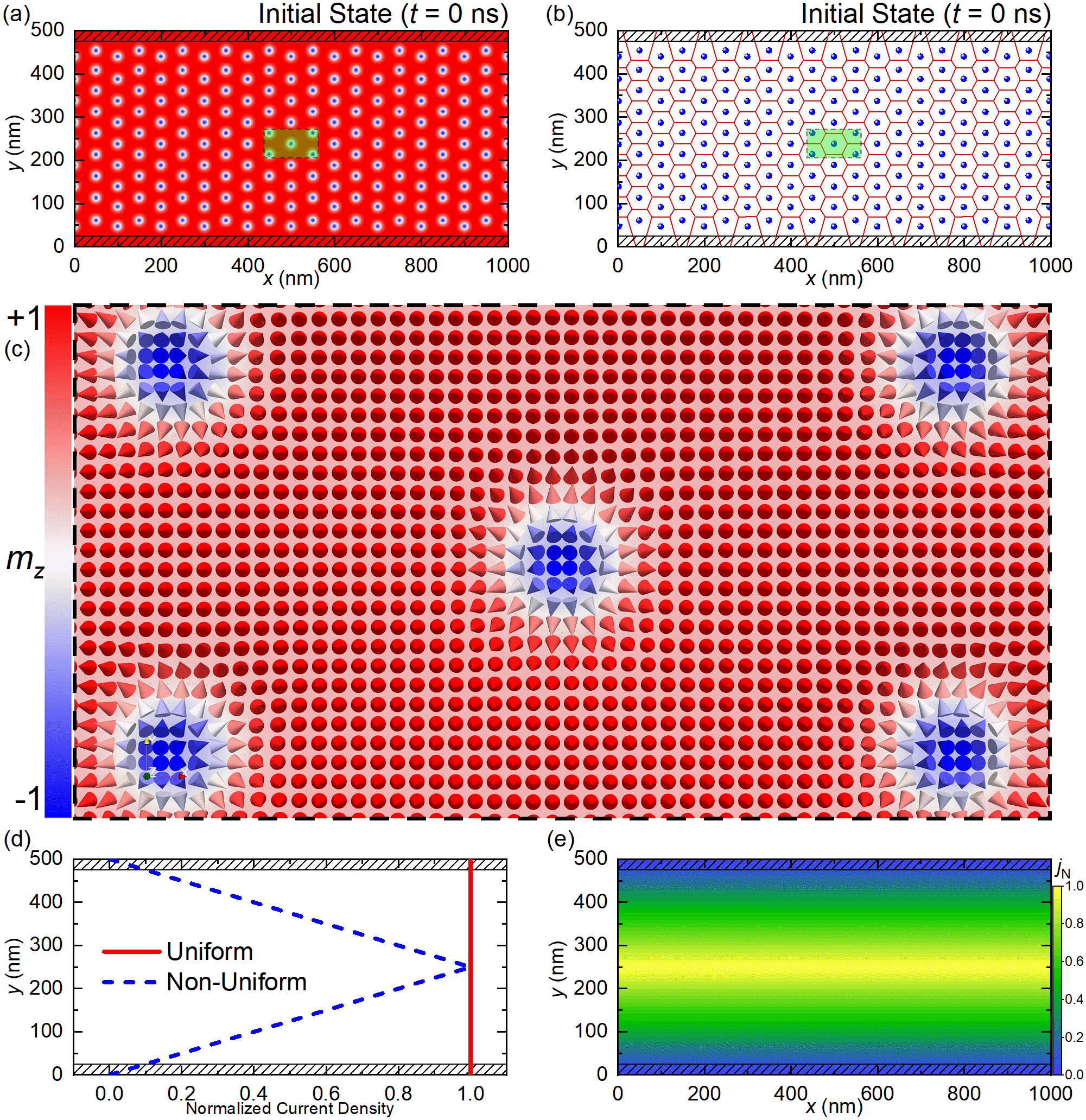}}
\caption{%
The initial state and driving current profiles.
(a) Top view of the initial state at $t=0$ ns. The upper and lower pipe edges are indicated by black-line patterns.
(b) Voronoi cell construction showing the triangular lattice structure and skyrmion locations in (a). The skyrmion center is indicated by the blue dot.
(c) Close-up top view of the skyrmion texture as indicated by the green boxes in (a) and (b). Each cone represents a spin. The color scale represents the $m_z$ component.
(d) Normalized current densities of the uniform and non-uniform currents along the lateral direction of the pipe.
(e) Normalized current density of the non-uniform current in the $x$-$y$ plane.
}
\label{FIG1}
\end{figure*}

\begin{figure*}[t]
\centerline{\includegraphics[width=0.50\textwidth]{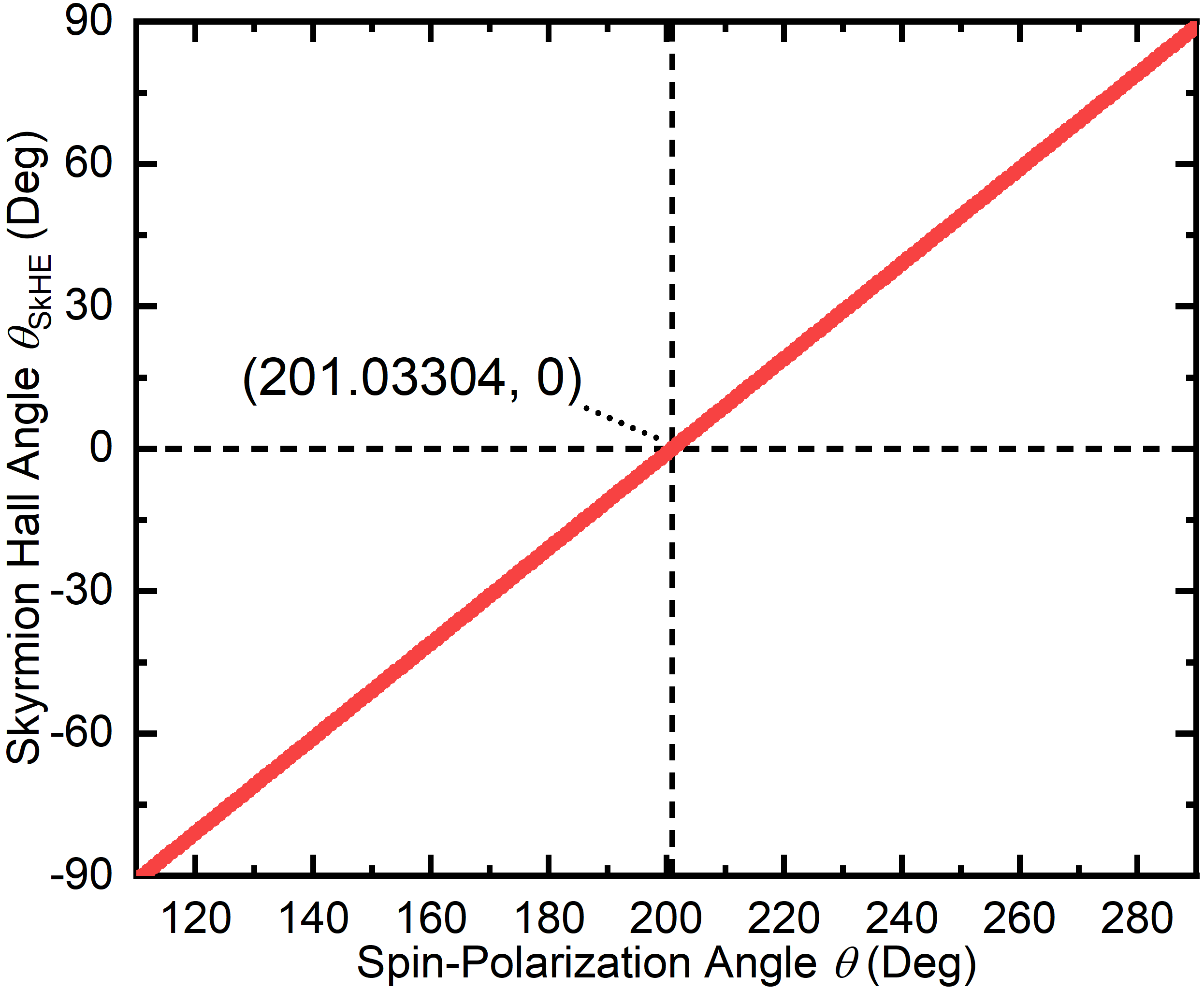}}
\caption{%
The intrinsic skyrmion Hall angle $\theta_{\text{SH}}$ of a single isolated skyrmion as a function of the spin-polarization angle $\theta$. In our studied system, the skyrmion moves toward the $+x$ direction (i.e., $\theta_{\text{SkHE}}=0^{\circ}$) when $\theta=201.03304^{\circ}$.
}
\label{FIG2}
\end{figure*}

\begin{figure*}[t]
\centerline{\includegraphics[width=0.99\textwidth]{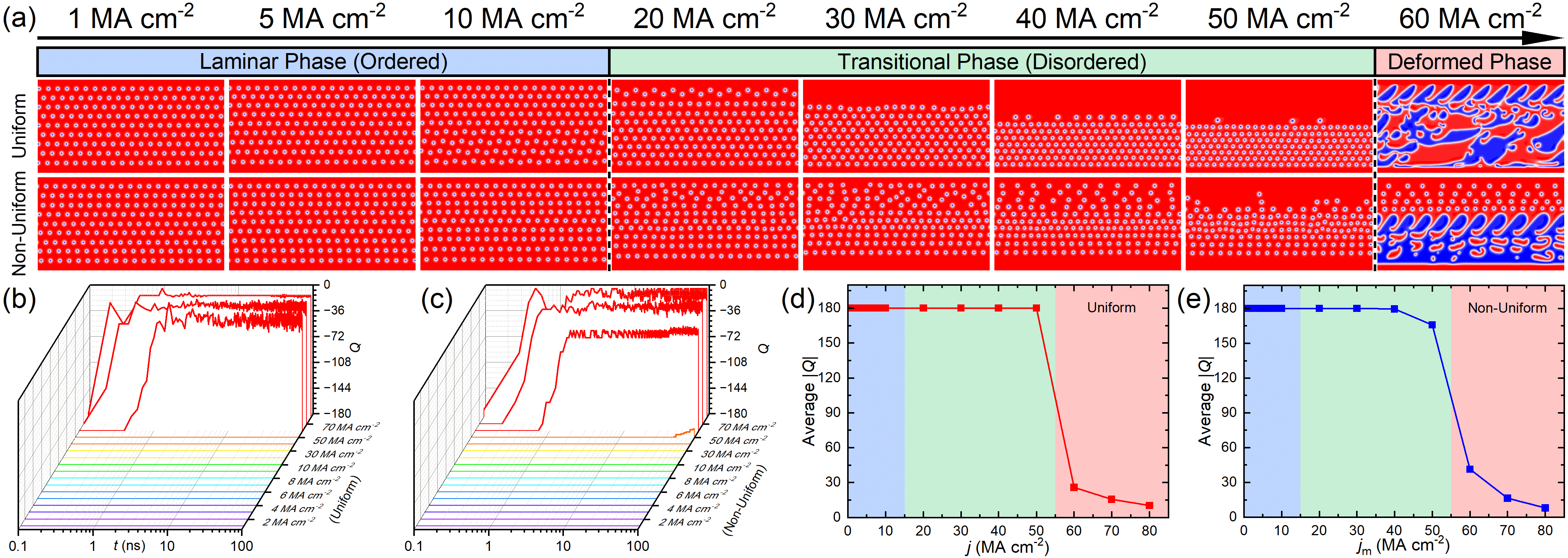}}
\caption{%
The dynamic phase diagram.
(a) Top views of the system at $t=500$ ns driven by a uniform or non-uniform current for a range of current densities, i.e., $j=j_{\text{m}}=1-60$ MA cm$^{-2}$, which indicate three different dynamic phases.
(b) Time-dependent $Q$ for the system driven by a uniform current of $j=1-80$ MA cm$^{-2}$.
(c) Time-dependent $Q$ for the system driven by a non-uniform current of $j_{\text{m}}=1-80$ MA cm$^{-2}$.
(d) $j$-dependent $|Q|$ averaged for $500$ ns for the system driven by a uniform current.
(e) $j_{\text{m}}$-dependent $|Q|$ averaged for $500$ ns for the system driven by a non-uniform current.
}
\label{FIG3}
\end{figure*}

\begin{figure*}[t]
\centerline{\includegraphics[width=0.95\textwidth]{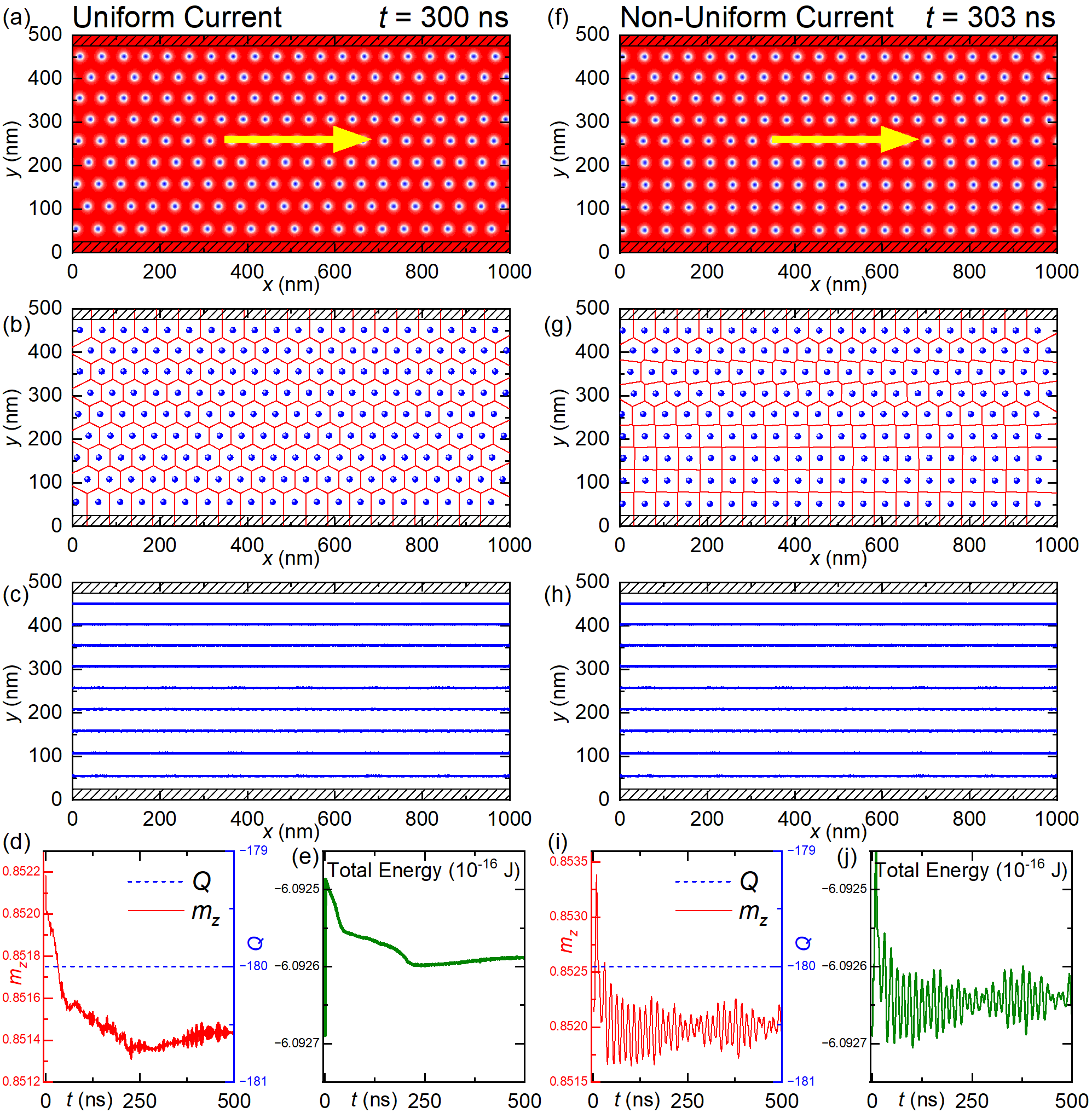}}
\caption{%
A laminar pipe flow of skyrmions driven by a small uniform or non-uniform current.
(a) Top view of the system driven by a uniform current ($j=3$ MA cm$^{-2}$) at $t=300$ ns. The yellow arrow indicates the skyrmion flowing direction.
(b) Voronoi cell construction of the state given in (a).
(c) Overlay of moving skyrmions driven by a uniform current in the pipe during $t=300-500$ ns, showing nine parallel pathlines.
(d) Time-dependent $m_z$, $Q$, and (e) total energy for the system driven by a uniform current.
(f) Top view of the system driven by a non-uniform current ($j_{\text{m}}=3$ MA cm$^{-2}$) at $t=303$ ns.
(g) Voronoi cell construction of the state given in (f).
(h) Overlay of moving skyrmions driven by a non-uniform current in the pipe during $t=300-500$ ns, showing nine parallel pathlines.
The skyrmions near the upper and lower pipe edges move much slower than that in the middle of the pipe as the skyrmion speed is proportional to the current density.
(i) Time-dependent $m_z$, $Q$, and (j) total energy for the system driven by a non-uniform current.
}
\label{FIG4}
\end{figure*}

\begin{figure*}[t]
\centerline{\includegraphics[width=0.50\textwidth]{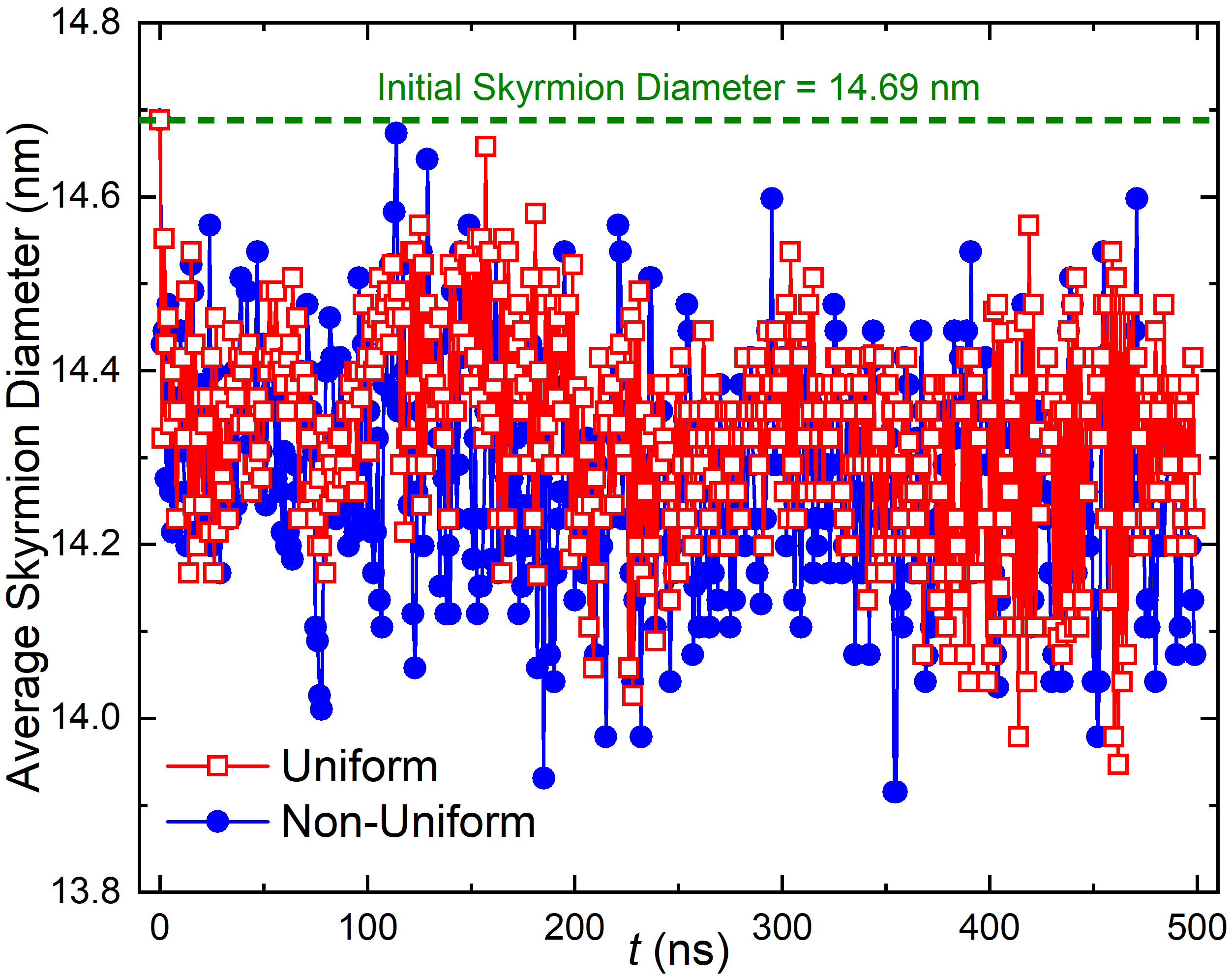}}
\caption{%
The average skyrmion diameter in the system driven by a small uniform current ($j=3$ MA cm$^{-2}$) or a non-uniform current ($j_{\text{m}}=3$ MA cm$^{-2}$) as a function of time. The initial skyrmion diameter equals $14.69$ nm before the application of the current.
}
\label{FIG5}
\end{figure*}

\begin{figure*}[t]
\centerline{\includegraphics[width=0.99\textwidth]{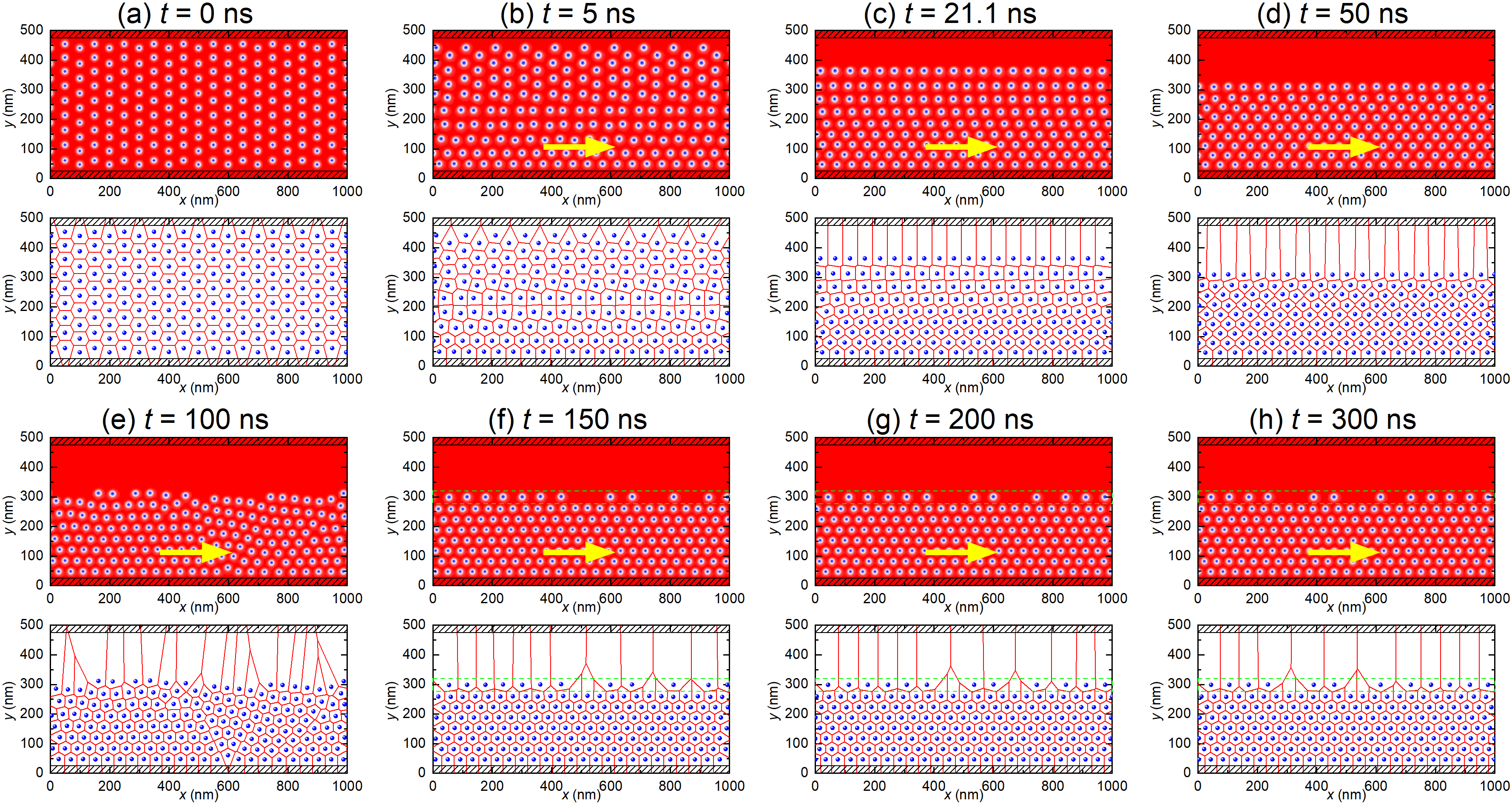}}
\caption{%
Transition from a pipe flow to an open-channel flow driven by a moderately large uniform current.
Top views and corresponding Voronoi cell constructions of the system driven by a uniform current ($j=40$ MA cm$^{-2}$) at (a) $t=0$ ns, (b) $t=5$ ns, (c) $t=21.1$ ns, (d) $t=50$ ns, (e) $t=100$ ns, (f) $t=150$ ns, (g) $t=200$ ns, and (h) $t=300$ ns.
}
\label{FIG6}
\end{figure*}

\begin{figure*}[t]
\centerline{\includegraphics[width=0.70\textwidth]{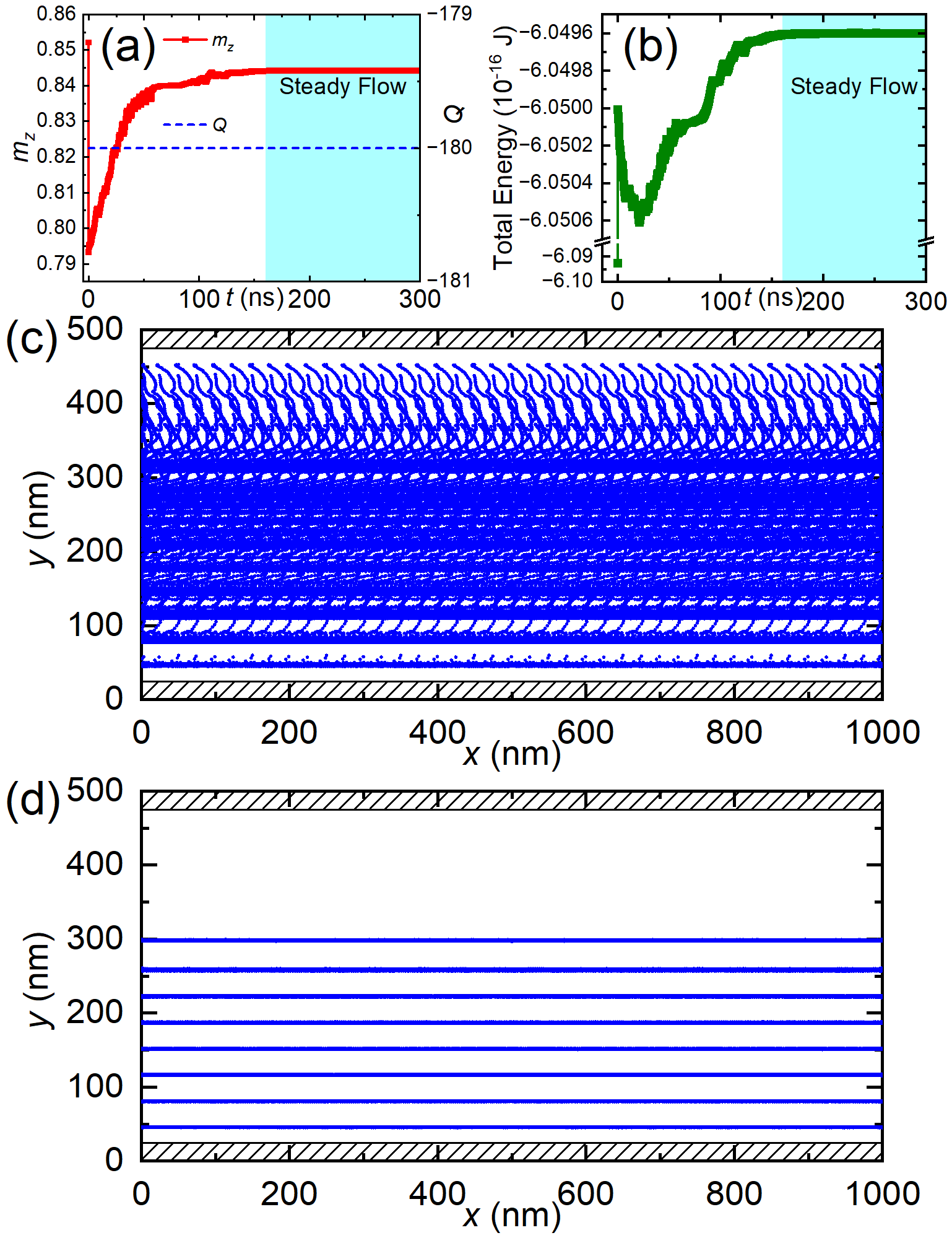}}
\caption{%
Transition from a pipe flow to an open-channel flow driven by a moderately large uniform current ($j=40$ MA cm$^{-2}$).
(a) Time-dependent $m_z$, $Q$, and (b) total energy for the system.
(c) Overlay of moving skyrmions during $t=0-50$ ns, showing both the lateral and longitudinal motion of skyrmions in the pipe.
(d) Overlay of moving skyrmions during $t=160-300$ ns, showing eight compressed parallel pathlines. The speed of skyrmions flowing along the pathline near $y=300$ nm is faster than that of the main skyrmion flow, forming a single shear layer of skyrmions at the surface of the main skyrmion flow.
}
\label{FIG7}
\end{figure*}

\begin{figure*}[t]
\centerline{\includegraphics[width=0.50\textwidth]{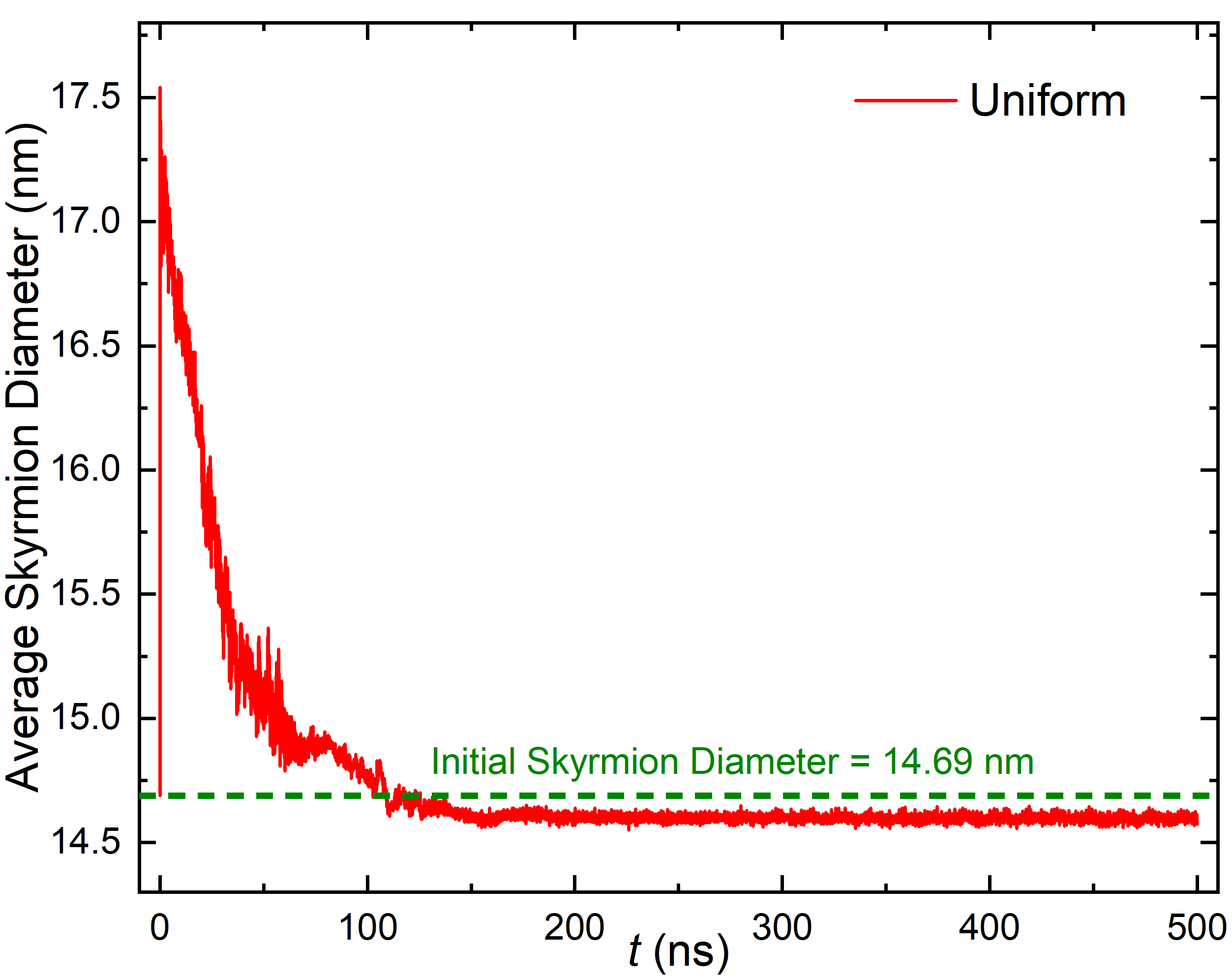}}
\caption{%
The average skyrmion diameter in the system driven by a moderately large uniform current ($j=40$ MA cm$^{-2}$) as a function of time. The initial skyrmion diameter equals $14.69$ nm before the application of the current.
}
\label{FIG8}
\end{figure*}

\begin{figure*}[t]
\centerline{\includegraphics[width=0.99\textwidth]{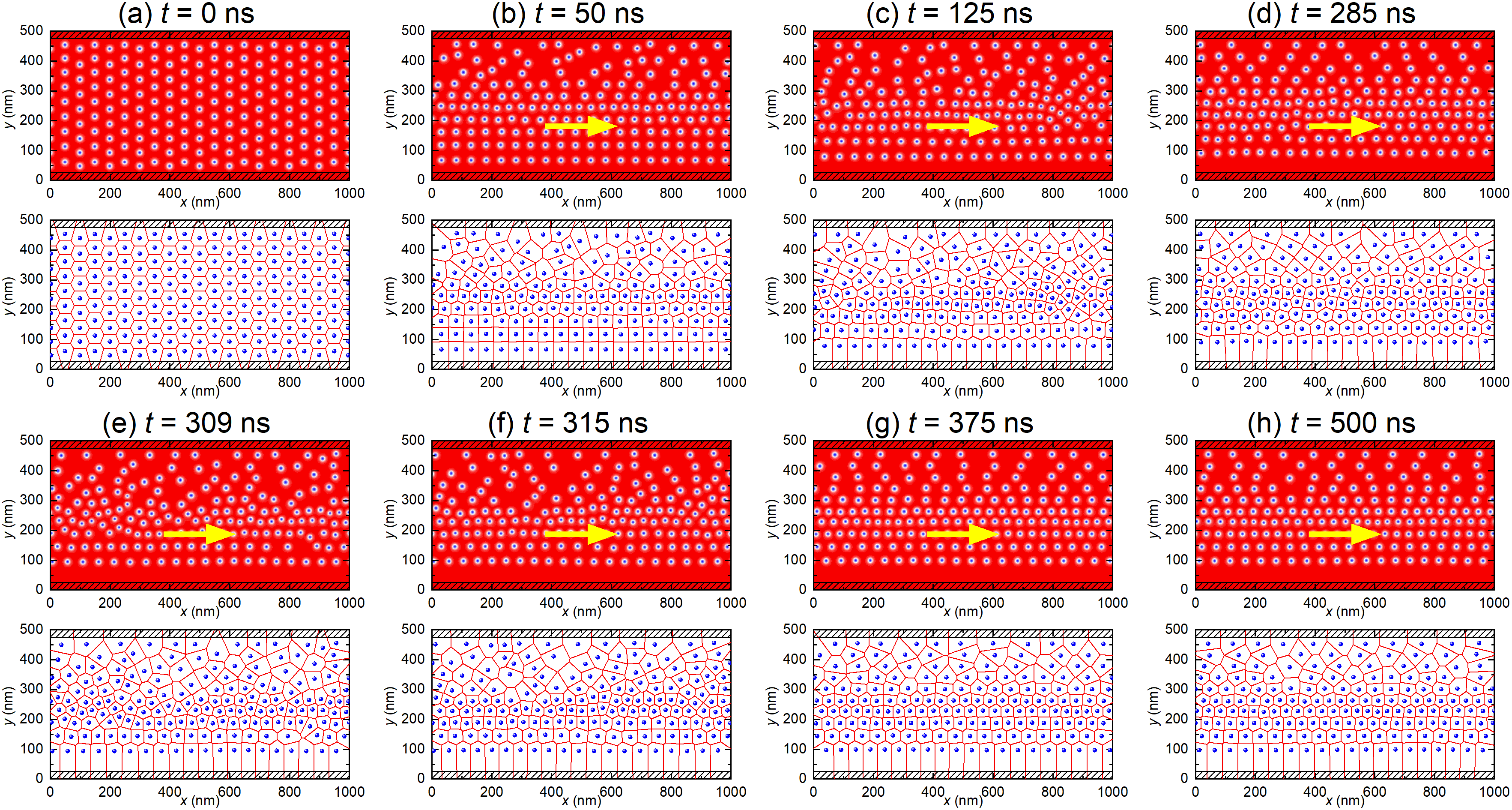}}
\caption{%
A transitional pipe flow of skyrmions driven by a moderately large non-uniform current.
Top views and corresponding Voronoi cell constructions of the system driven by a non-uniform current ($j_{\text{m}}=40$ MA cm$^{-2}$) at (a) $t=0$ ns, (b) $t=50$ ns, (c) $t=125$ ns, (d) $t=285$ ns, (e) $t=309$ ns, (f) $t=315$ ns, (g) $t=375$ ns, and (h) $t=500$ ns.
}
\label{FIG9}
\end{figure*}

\begin{figure*}[t]
\centerline{\includegraphics[width=0.90\textwidth]{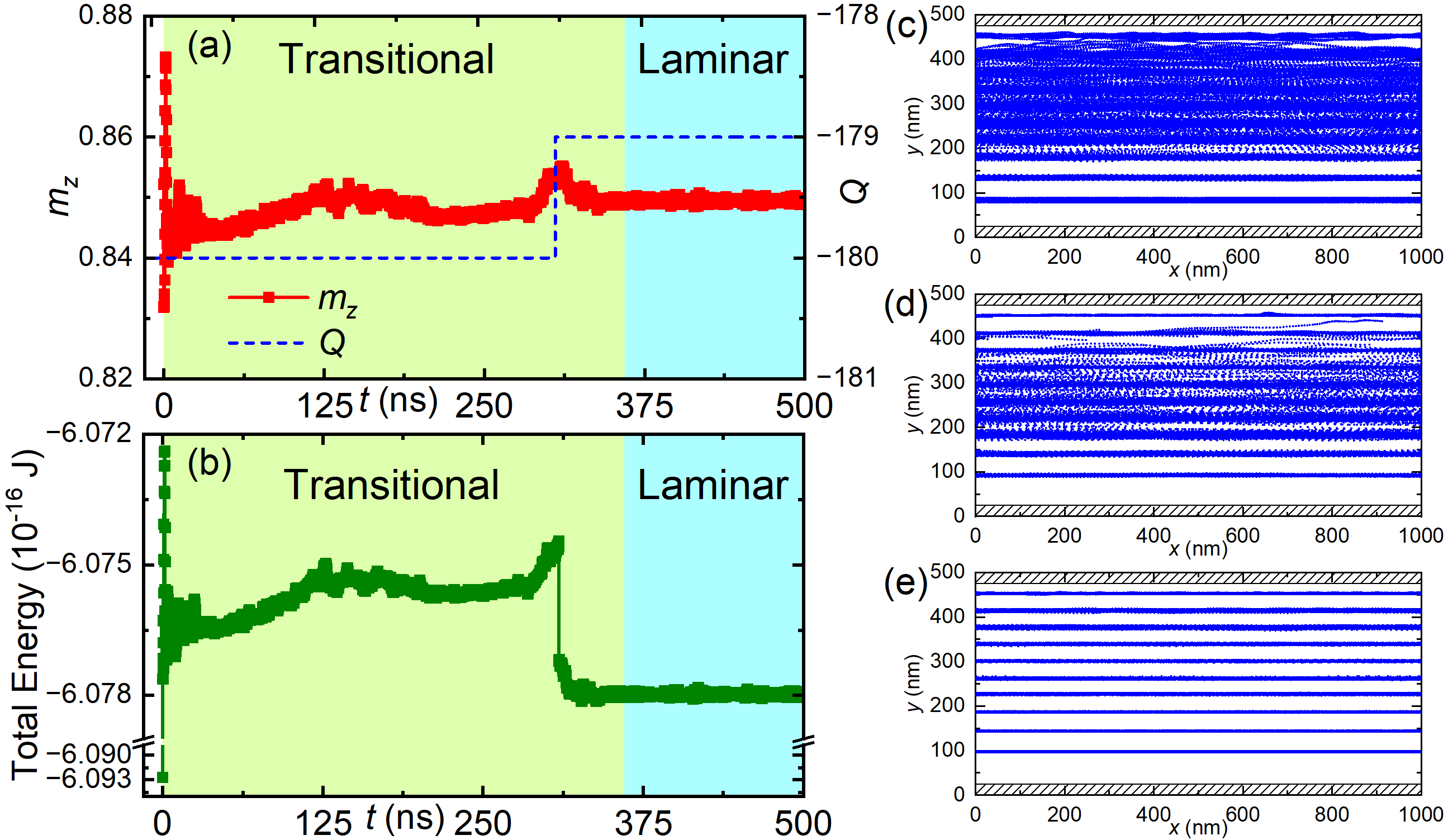}}
\caption{%
A transitional pipe flow of skyrmions driven by a moderately large non-uniform current ($j_{\text{m}}=40$ MA cm$^{-2}$).
(a) Time-dependent $m_z$, $Q$, and (b) total energy for the system.
Overlays of moving skyrmions during (c) $t=125-175$ ns and (d) $t=285-309$ ns show some transiently disordered skyrmion behaviors in the upper half of the pipe.
(e) Overlay of moving skyrmions during $t=360-500$ ns, showing $10$ parallel pathlines, where the skyrmions flow in a laminar fashion.
}
\label{FIG10}
\end{figure*}

\begin{figure*}[t]
\centerline{\includegraphics[width=0.50\textwidth]{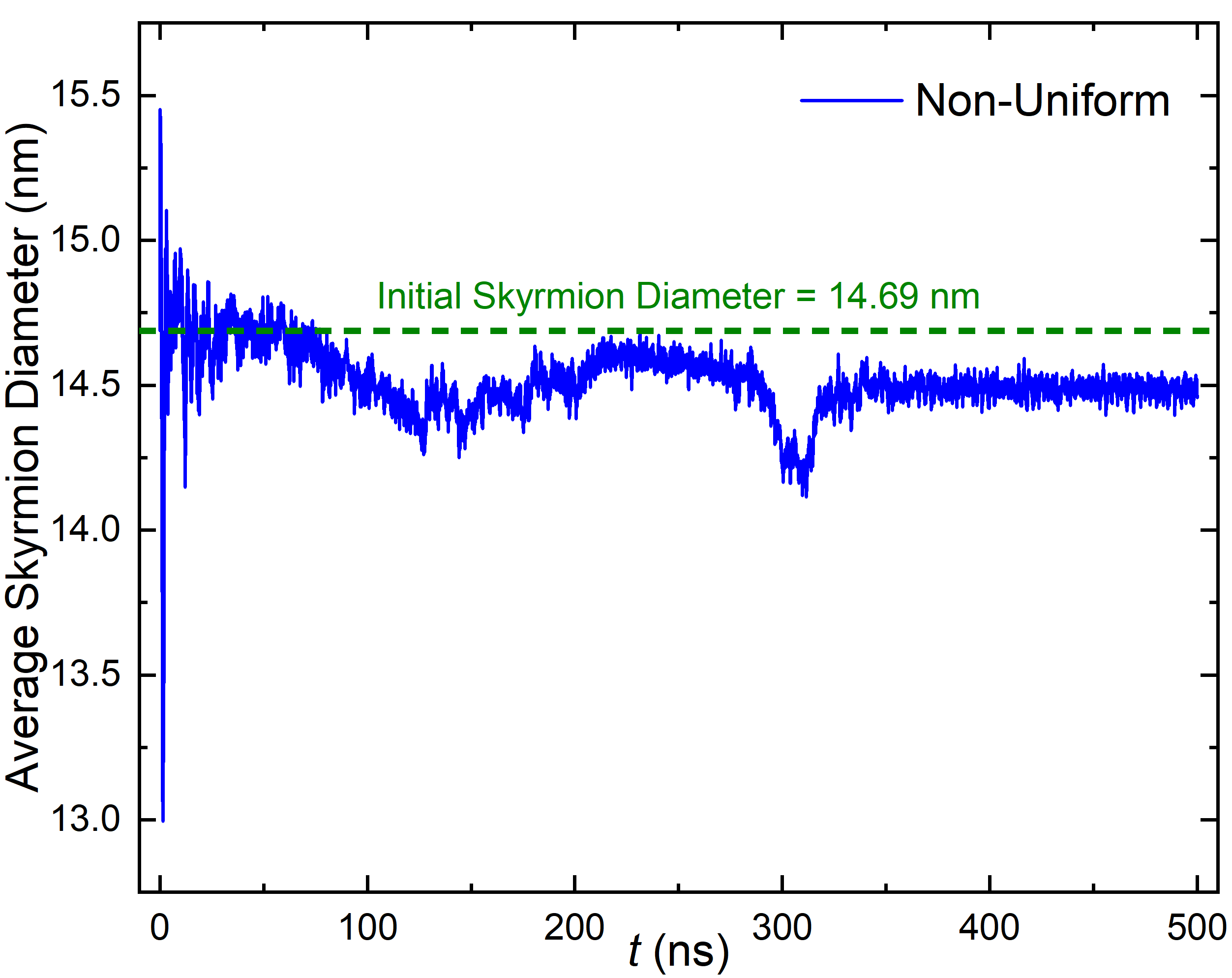}}
\caption{%
The average skyrmion diameter in the system driven by a moderately large non-uniform current ($j_{\text{m}}=40$ MA cm$^{-2}$) as a function of time. The initial skyrmion diameter equals $14.69$ nm before the application of the current.
}
\label{FIG11}
\end{figure*}

\begin{figure*}[t]
\centerline{\includegraphics[width=0.99\textwidth]{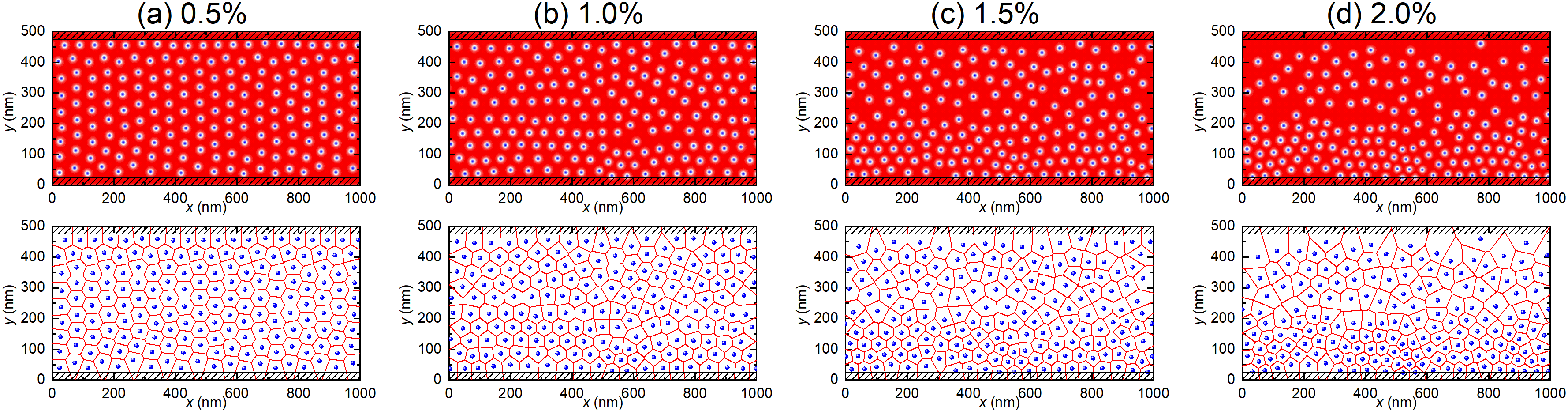}}
\caption{%
Effect of magnetic anisotropy variation on the skyrmion pipe flow driven by a small uniform current ($j=3$ MA cm$^{-2}$).
Top views and corresponding Voronoi cell constructions at $t=300$ ns for the systems with anisotropy variations of (a) $0.5\%$, (b) $1.0\%$, (c) $1.5\%$, and (d) $2.0\%$.
}
\label{FIG12}
\end{figure*}

\begin{figure*}[t]
\centerline{\includegraphics[width=0.85\textwidth]{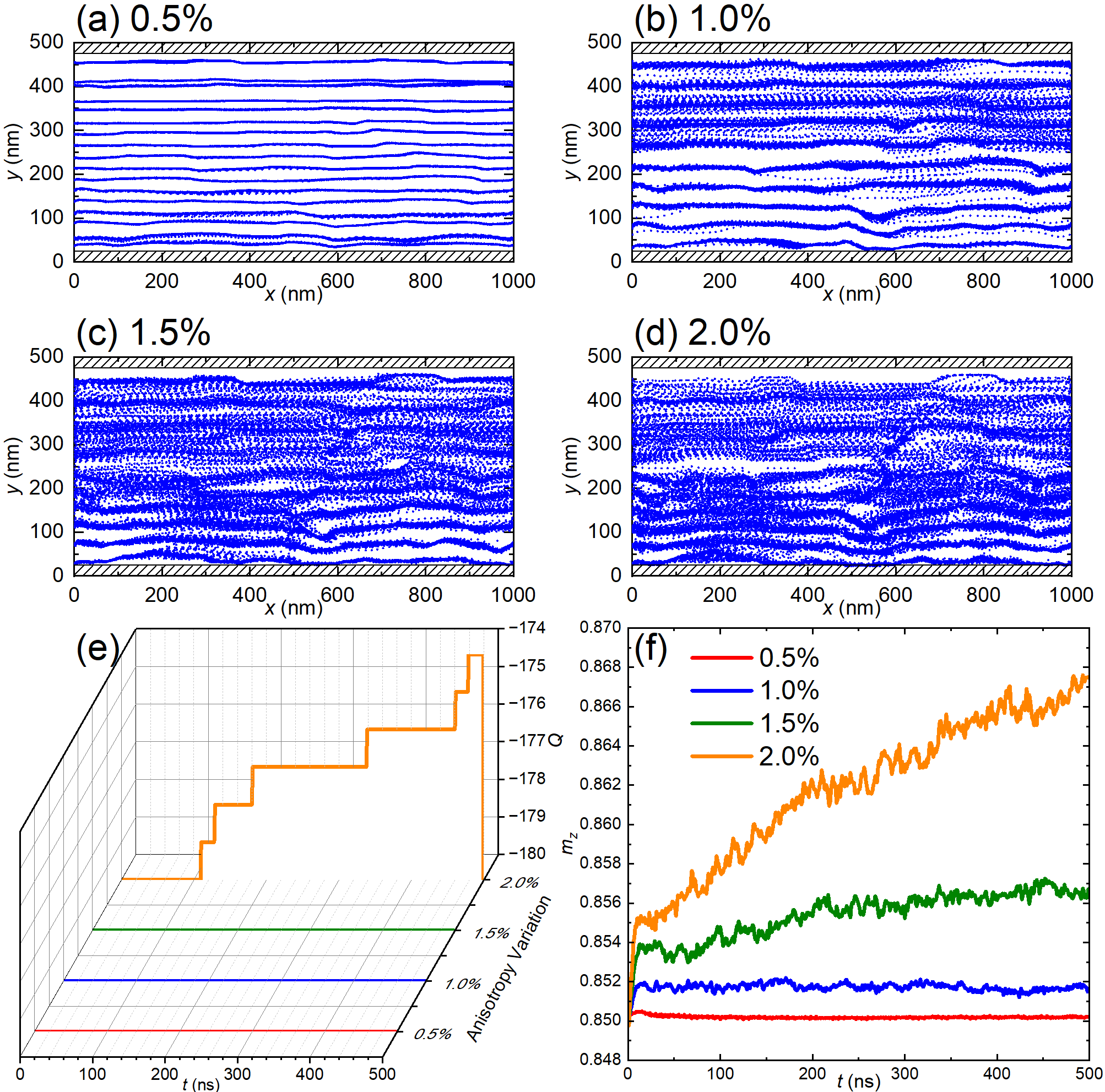}}
\caption{%
Effect of magnetic anisotropy variation on the skyrmion pipe flow driven by a small uniform current ($j=3$ MA cm$^{-2}$).
Overlay of moving skyrmions driven by a small uniform current during $t=300-500$ ns in the system with an anisotropy variation of (a) $0.5\%$, (b) $1.0\%$, (c) $1.5\%$, and (d) $2.0\%$, which shows multiple wavy pathlines.
(e) Time-dependent $Q$ and (f) $m_z$ for the systems with different anisotropy variations.
}
\label{FIG13}
\end{figure*}

\begin{figure*}[t]
\centerline{\includegraphics[width=0.50\textwidth]{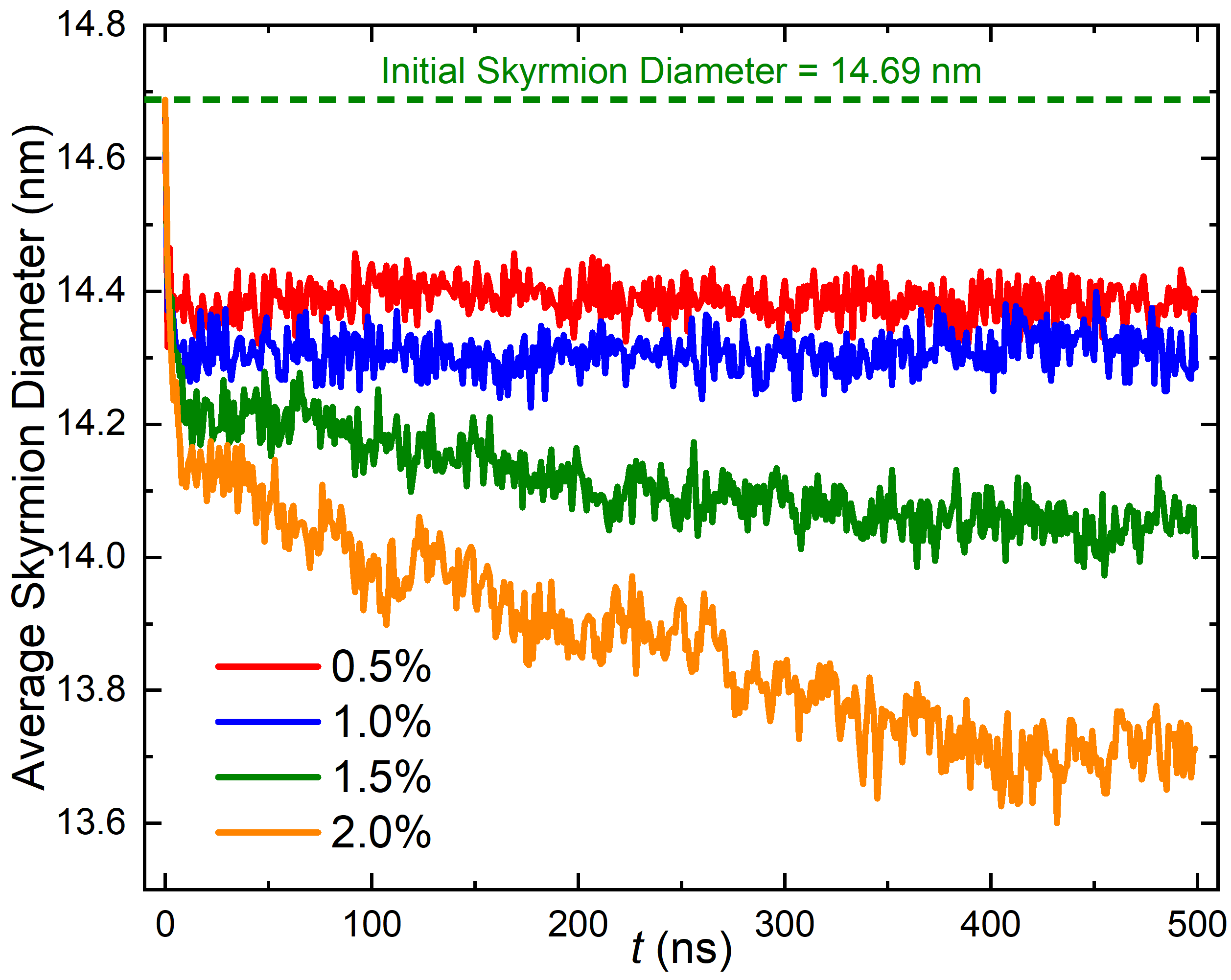}}
\caption{%
The average skyrmion diameter as a function of time in the system with magnetic anisotropy variation driven by a small uniform current ($j=3$ MA cm$^{-2}$). The initial skyrmion diameter equals $14.69$ nm in the system with no anisotropy variation before the application of the current.
}
\label{FIG14}
\end{figure*}

\begin{figure*}[t]
\centerline{\includegraphics[width=0.99\textwidth]{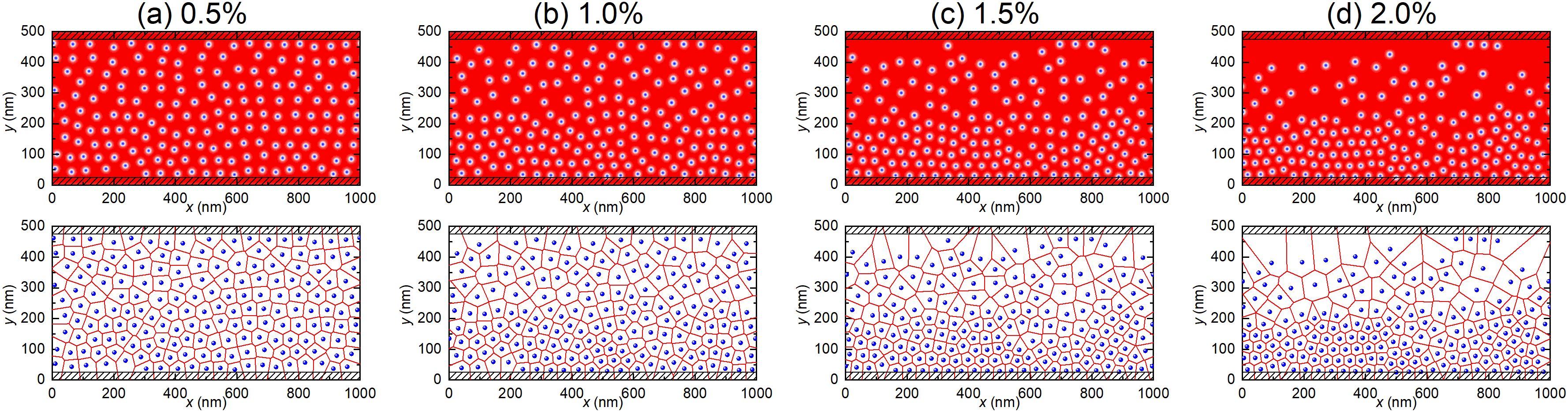}}
\caption{%
Effect of magnetic anisotropy variation on the skyrmion pipe flow driven by a small non-uniform current ($j_{\text{m}}=3$ MA cm$^{-2}$).
Top views and corresponding Voronoi cell constructions at $t=303$ ns for the systems with anisotropy variations of (a) $0.5\%$, (b) $1.0\%$, (c) $1.5\%$, and (d) $2.0\%$.
}
\label{FIG15}
\end{figure*}

\begin{figure*}[t]
\centerline{\includegraphics[width=0.85\textwidth]{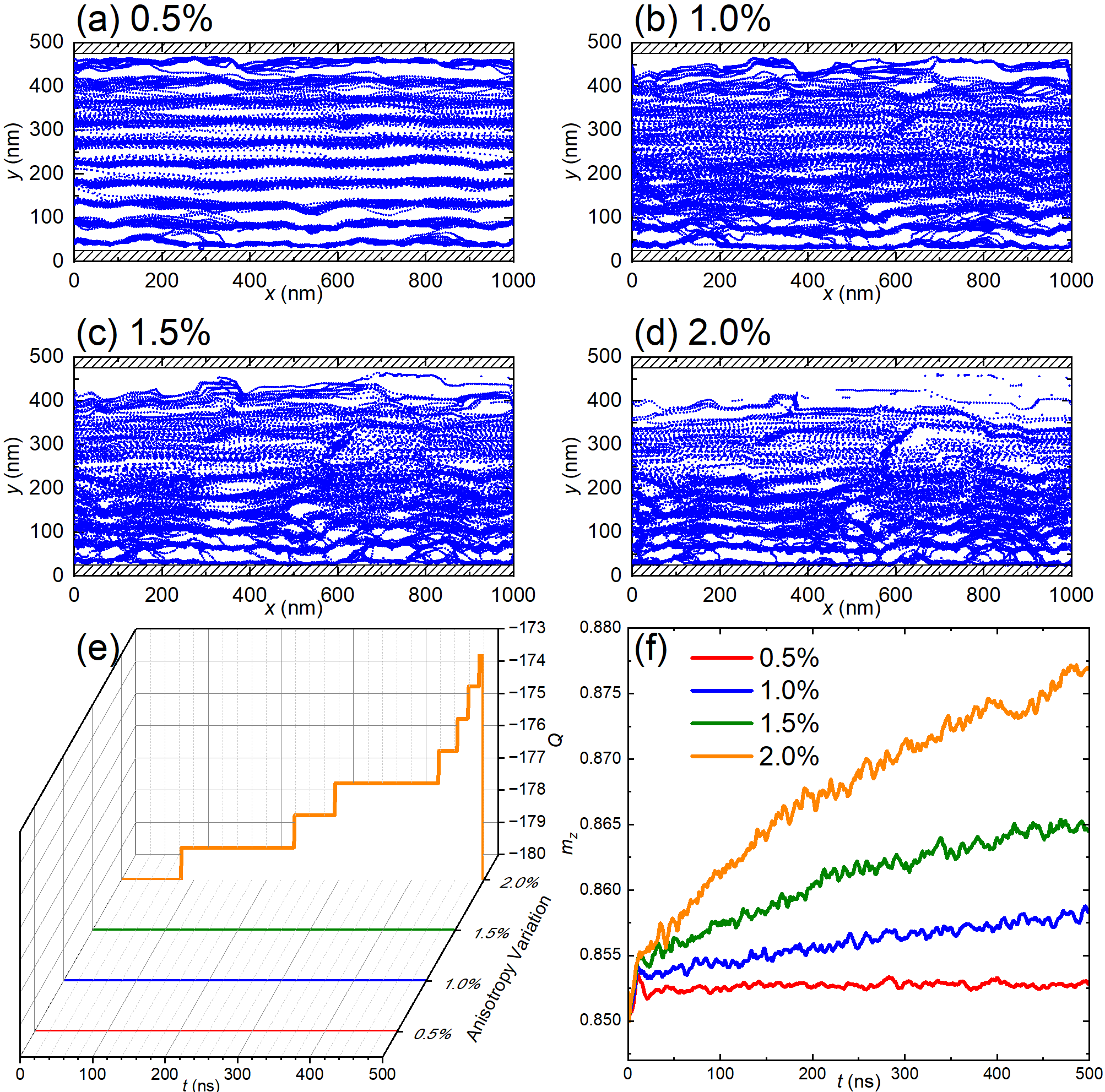}}
\caption{%
Effect of magnetic anisotropy variation on the skyrmion pipe flow driven by a small non-uniform current ($j_{\text{m}}=3$ MA cm$^{-2}$).
Overlay of moving skyrmions driven by a small non-uniform current during $t=300-500$ ns in the system with an anisotropy variation of (a) $0.5\%$, (b) $1.0\%$, (c) $1.5\%$, and (d) $2.0\%$, which shows multiple wavy pathlines.
(e) Time-dependent $Q$ and (f) $m_z$ for the systems with different anisotropy variations.
}
\label{FIG16}
\end{figure*}

\begin{figure*}[t]
\centerline{\includegraphics[width=0.50\textwidth]{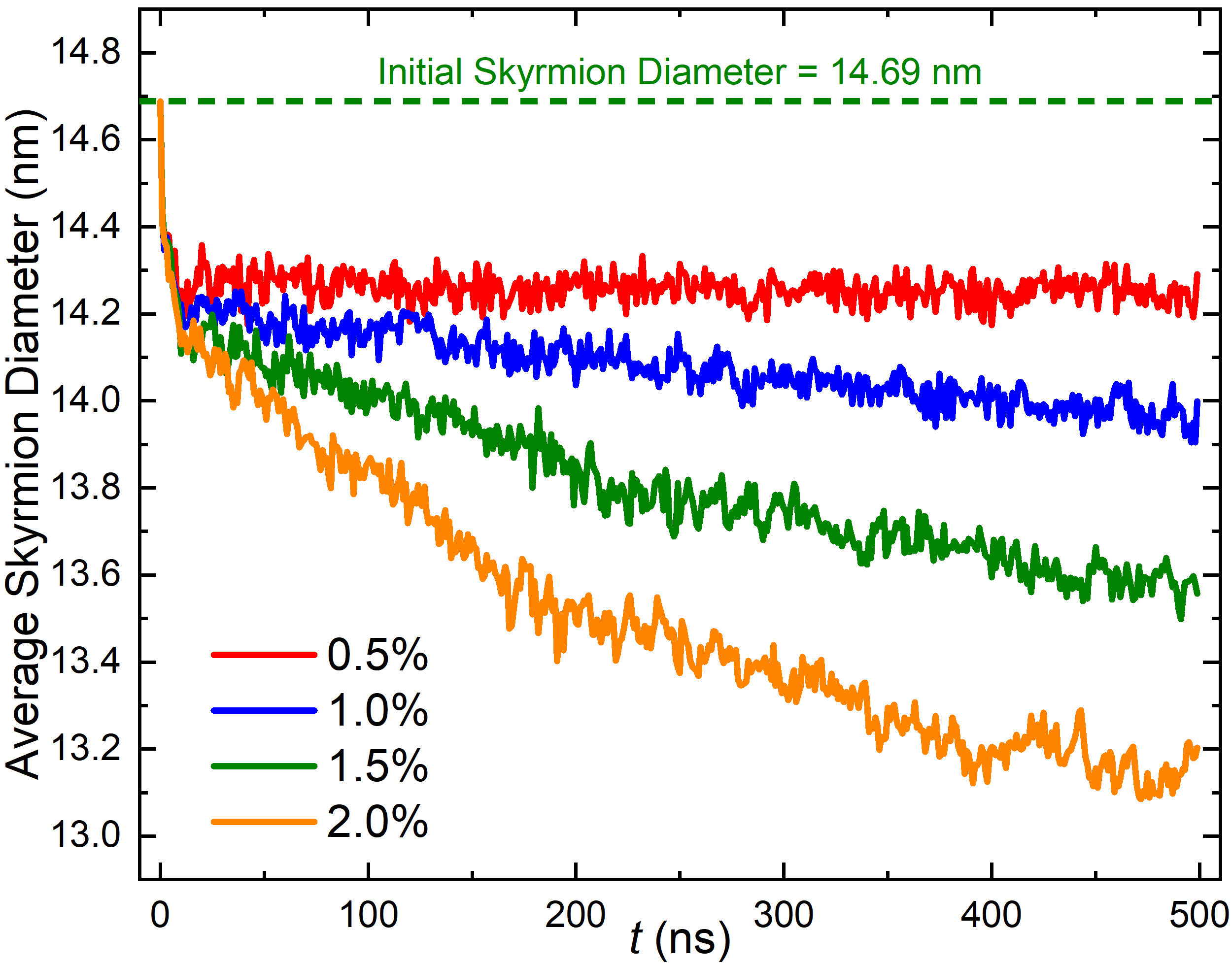}}
\caption{%
The average skyrmion diameter as a function of time in the system with magnetic anisotropy variation driven by a small non-uniform current ($j_{\text{m}}=3$ MA cm$^{-2}$). The initial skyrmion diameter equals $14.69$ nm in the system with no anisotropy variation before the application of the current.
}
\label{FIG17}
\end{figure*}

The advance of hardware and software technologies will benefit future information-driven society~\cite{Ayyash_IEEE2015}.
The essential information-processing hardware include storage, computing, and communication devices~\cite{Galindo_RMP2002,Goda_PIEEE2012}.
Particularly, the performance of storage and computing hardware may be improved by replacing conventional information carriers with topologically-nontrivial ones~\cite{Nagaosa_NNANO2013,Mochizuki_JPCM2015,Wiesendanger_NATREVMAT2016,Finocchio_JPD2016,Fert_NATREVMAT2017,Wanjun_PHYSREP2017,Everschor_JAP2018,Zhang_JPCM2020,Back_JPD2020,Gobel_PHYSREP2021,Yu_JMMM2021,Kanazawa_AM2017,Fujishiro_APL2020,Tokura_CR2021,Del-Valle_2022}.
A promising candidate is the magnetic skyrmion~\cite{Bogdanov_SPJETP1989,Roszler_NATURE2006,Muhlbauer_SCIENCE2009,Yu_NATURE2010,Bogdanov_NRP2020}, which is a particle-like object~\cite{Reichhardt_RMP2022,Lin_PRB2013,Reichhardt_PRL2015,Reichhardt_PRB2015A,Reichhardt_JPCM2019,Reichhardt_PRB2019,Reichhardt_PRB2020,Souza_2023B,Muratov_2022,Reichhardt_PRB2015B,Reichhardt_NJP2015,Reichhardt_PRB2016,Reichhardt_NJP2016,Reichhardt_PRL2018,Vizarim_PRB2020,Vizarim_PRB2022,Reichhardt_PRB2022} and can be controlled effectively by spin currents~\cite{Zang_PRL2011,Wanjun_SCIENCE2015,Wanjun_NPHYS2017,Litzius_NPHYS2017,Sampaio_NN2013,Tomasello_SREP2014,Xichao_PRB2016B,Xichao_PRB2022A,Xichao_PRB2022B} and electric fields~\cite{Zhang_SREP2015,Schott_NL2017,Ma_NL2019,Liu_PRA2019,Bhattacharya_NE2020,Yang_AM2023}.
The use of skyrmions in hardware could, in principle, enhance the radiation tolerance and reduce the energy consumption~\cite{Kang_PIEEE2016,Li_MH2021,Luo_APLM2021,Marrows_APL2021,Vakili_JAP2021}, which may make it possible to operate the hardware efficiently under extreme conditions.
Besides, the magnetic skyrmions with dimensions down to a few nanometers may also be used as building blocks for quantum versions of classical information-processing hardware~\cite{Lohani_PRX2019,Psaroudaki_PRL2021,Psaroudaki_PRB2022,Xia_PRL2022}.

It is therefore important to explore the complex dynamics of skyrmions in nanoscale device geometries in order to realize skyrmion-based functional devices.
The single and collective skyrmion dynamics are fundamental for the manipulation of skyrmionic bits in magnetic substrates~\cite{Nagaosa_NNANO2013,Mochizuki_JPCM2015,Wiesendanger_NATREVMAT2016,Finocchio_JPD2016,Fert_NATREVMAT2017,Wanjun_PHYSREP2017,Everschor_JAP2018,Zhang_JPCM2020,Back_JPD2020,Gobel_PHYSREP2021,Reichhardt_RMP2022,Kang_PIEEE2016,Li_MH2021,Luo_APLM2021,Marrows_APL2021,Vakili_JAP2021,Yu_JMMM2021,Kanazawa_AM2017,Fujishiro_APL2020,Tokura_CR2021,Del-Valle_2022}.
A recent comprehensive review~\cite{Reichhardt_RMP2022} has highlighted the dynamic properties of skyrmions interacting with disorder and nanostructures, which are believed to be important for technological applications.
However, the control of skyrmion flow dynamics could be a challenging task due to the complex skyrmion-skyrmion and skyrmion-substrate interactions~\cite{Reichhardt_NC2020,Yu_NC2012,Okuyama_CP2019,Sato_PRB2019,Souza_PRB2021,Souza_NJP2022,Souza_2023,Xichao_PRB2022A,Xichao_PRB2022B}.

An important problem of the flow dynamics of fluid particles in pipes is whether it is described by the laminar, turbulent, or transitional dynamics~\cite{White_2011,Eckhardt_ARFM2007,Guha_ARFM2008,Zahtila_JFM2003,Barkley_PRL2005,Miguel_PRL2011,Squires_RMP2005,Knebel_NP2016}.
Skyrmions are quasiparticles and can interact with each other in a repulsive manner~\cite{Reichhardt_RMP2022,Lin_PRB2013}.
Thus, a large amount of nanoscale skyrmions may also form a very special type of plastic or viscous fluid flow~\cite{Reichhardt_NC2020,Yu_NC2012,Okuyama_CP2019,Sato_PRB2019,Souza_PRB2021,Souza_NJP2022,Souza_2023,Xichao_PRB2022A,Xichao_PRB2022B,Reichhardt_JPCM2019,Reichhardt_PRB2019,Reichhardt_PRB2020} in pipe channels and show typical fluid dynamics.
The transition between different dynamic phases of skyrmion flows could result in complex and interesting transitional transport phenomena, which are of fundamental physical interest and may play an important role in practical applications based on the transport of skyrmionic quasiparticles.
In this work, we report the laminar and transiently disordered behaviors of a skyrmion flow in a two-dimensional (2D) pipe, with a focus on the dynamics in the absence of pinning and defect effects.

\section{Theoretical model and simulations}
\label{se:Methods}

We simulate the current-driven dynamics of many compact N{\'e}el-type skyrmions in a 2D ferromagnetic pipe channel with the interfacial Dzyaloshinskii-Moriya (DM) interaction~\cite{Dzyaloshinsky,Moriya}.
The length, width, and thickness of the model are equal to $1000$ nm, $500$ nm, and $1$ nm, respectively.
The periodic boundary condition is only applied in the $x$ direction, while $10$ edge spins on the upper and lower edges of the pipe are assumed to have enhanced perpendicular magnetic anisotropy (PMA)~\cite{Juge_NL2021,Ohara_NL2021,Zhang_CP2021}, which create a confined pipe channel geometry [Figs.~\ref{FIG1}(a)~and~\ref{FIG1}(b)].
The spin dynamics is controlled by the Landau-Lifshitz-Gilbert (LLG) equation augmented with the damping-like spin-orbit torque~\cite{Sampaio_NN2013,Tomasello_SREP2014,Xichao_PRB2016B,Xichao_PRB2022A,Xichao_PRB2022B}, which can be generated by the spin Hall effect in a heavy-metal substrate~\cite{Tomasello_SREP2014,Wanjun_SCIENCE2015,Wanjun_NPHYS2017,Litzius_NPHYS2017,Sinova_RMP2015}.
The field-like torque is not considered as it does not drive the dynamics of a compact and rigid skyrmion~\cite{Tomasello_SREP2014}.
Thus, the spin dynamics equation is
\begin{equation}
\label{eq:LLGS-SOT-Damping-Like}
\partial_{t}\boldsymbol{m}=-\gamma_{0}\boldsymbol{m}\times\boldsymbol{h}_{\text{eff}}+\alpha(\boldsymbol{m}\times\partial_{t}\boldsymbol{m})+\boldsymbol{\tau}_{\text{d}},
\end{equation}
where $\boldsymbol{m}$ is the reduced magnetization,
$t$ is the time,
$\gamma_0$ is the absolute gyromagnetic ratio,
$\alpha$ is the Gilbert damping parameter,
$\boldsymbol{h}_{\rm{eff}}=-\frac{1}{\mu_{0}M_{\text{S}}}\cdot\frac{\delta\varepsilon}{\delta\boldsymbol{m}}$ is the effective field.
$\mu_{0}$, $M_{\text{S}}$, and $\varepsilon$ denote the vacuum permeability constant, saturation magnetization, and average energy density, respectively.
The system energy terms include the exchange energy, DM interaction energy, PMA energy, and demagnetization energy, as expressed in the average energy density below~\cite{Sampaio_NN2013,Tomasello_SREP2014,Xichao_PRB2016B,Xichao_PRB2022A,Xichao_PRB2022B}
\begin{equation}
\label{eq:energy-density}
\begin{split}
\varepsilon=&A\left(\nabla\boldsymbol{m}\right)^{2}+D\left[m_{z}\left(\boldsymbol{m}\cdot\nabla\right)-\left(\nabla\cdot\boldsymbol{m}\right)m_{z}\right] \\
-&K(\boldsymbol{n}\cdot\boldsymbol{m})^2-\frac{M_{\text{S}}}{2}(\boldsymbol{m}\cdot\boldsymbol{B}_{\text{d}}),
\end{split}
\end{equation}
where $A$, $D$, and $K$ are the ferromagnetic exchange, DM interaction, and PMA constants, respectively. $\boldsymbol{B}_{\text{d}}$ is the demagnetization field. $\boldsymbol{n}$ is the unit surface normal vector. $m_z$ is the out-of-plane component of $\boldsymbol{m}$.
The damping-like torque
$\boldsymbol{\tau}_{\text{d}}=u\left(\boldsymbol{m}\times\boldsymbol{p}\times\boldsymbol{m}\right)$ with the coefficient being
$u=\left|\left(\gamma_{0}\hbar/\mu_{0}e\right)\right|\cdot\left(j\theta_{\text{SH}}/2aM_{\text{S}}\right)$.
$\hbar$ is the reduced Planck constant, $e$ is the electron charge, $a$ is the magnetic layer thickness, $j$ is the current density, $\theta_{\text{SH}}$ is the spin Hall angle, and $\boldsymbol{p}$ is the spin polarization direction.
The default parameters are~\cite{Sampaio_NN2013,Tomasello_SREP2014,Xichao_PRB2016B,Xichao_PRB2022A,Xichao_PRB2022B}: $\gamma_{0}=2.211\times 10^{5}$ m A$^{-1}$ s$^{-1}$, $\alpha=0.3$, $M_{\text{S}}=580$ kA m$^{-1}$, the exchange constant $A=15$ pJ m$^{-1}$, the PMA constant $K=0.8$ MJ m$^{-3}$, and the DM interaction constant $D=3$ mJ m$^{-2}$.
The driving force is solely controlled by the current density (i.e., $u\sim j$) with the assumption of $\theta_{\text{SH}}=1$.
All simulations are performed by the \textsc{mumax$^3$} micromagnetic simulator~\cite{MuMax} on an NVIDIA GeForce RTX 3060 Ti graphics processing unit.
The mesh size is set to $2.5$ $\times$ $2.5$ $\times$ $1$ nm$^3$ to ensure good computational accuracy and efficiency.

\section{Results and Discussion}
\label{se:Results}

\subsection{The initial skyrmion Hall angle}
\label{se:SkHA}

We focus on the skyrmion dynamics with an initially zero skyrmion Hall angle (i.e., the overdamped case~\cite{Reichhardt_RMP2022,Lin_PRB2013,Reichhardt_PRL2015,Reichhardt_PRB2015A,Reichhardt_JPCM2019,Reichhardt_PRB2019,Reichhardt_PRB2020,Reichhardt_PRB2021,Souza_2023B}), where skyrmions move in the $+x$ direction and can interact repulsively with pipe edges to form a pipe flow.
Hence, we set the intrinsic skyrmion Hall angle $\theta_{\text{SkHE}}$ to zero by changing the spin polarization direction $\boldsymbol{p}$~\cite{Xichao_PRB2022B}.
The spin-polarization angle between $\boldsymbol{p}$ and the $+x$ direction is defined as $\theta$, and $\theta_{\text{SkHE}}$ is defined as the angle between the skyrmion velocity $\boldsymbol{v}$ and the $+x$ direction~\cite{Wanjun_NPHYS2017}, i.e., $\theta_{\text{SkHE}}=\text{arctan}(v_y/v_x)$.
Note that the arctan function returns the inverse of the tangent function.
In experiments, $\theta$ could be controlled by tuning the in-plane electron flow direction $\boldsymbol{\hat{j}}_{e}$ in the heavy-metal substrate as $\boldsymbol{p}=\boldsymbol{\hat{j}}_{e}\times\boldsymbol{n}$ with $\boldsymbol{n}$ being the surface normal vector~\cite{Sinova_RMP2015,Tomasello_SREP2014,Sampaio_NN2013,Xichao_PRB2022B}.
Namely, an in-plane electron current is injected into the heavy-metal layer underneath the ferromagnetic layer, which results in a vertical spin current propagating into the ferromagnetic layer due to the spin Hall effect~\cite{Sinova_RMP2015,Tomasello_SREP2014,Sampaio_NN2013,Xichao_PRB2022B}, where the magnitude and spin-polarization angle of the spin current depend on the magnitude and direction of the in-plane electron current, respectively. The magnitude and direction of the injection electron current could be controlled by engineering the geometry of the heavy-metal layer. For example, a gradient electron current profile could be realized by fabricating a heavy-metal layer with a gradient thickness.

To determine $\theta$ for $\theta_{\text{SkHE}}=0^{\circ}$, we analyze $\theta_{\text{SkHE}}$ of a compact skyrmion using the Thiele equation~\cite{Thiele_PRL1973,Tomasello_SREP2014,Wang_PRB2019,Xichao_PRB2022B},
\begin{equation}
\boldsymbol{G}\times\boldsymbol{v}-\alpha\boldsymbol{\D}\cdot\boldsymbol{v}-4\pi\boldsymbol{\B}\cdot\boldsymbol{j}_{e}=\boldsymbol{0},
\label{eq:TME-CPP}
\end{equation}
where $\boldsymbol{v}=(v_x,v_y)$ is the steady skyrmion velocity, and
$\boldsymbol{j}_{e}=(j_x,j_y)=(-j_e\sin\theta,j_e\cos\theta)$ is the electron current.
$\boldsymbol{G}=(0,0,-4\pi Q)$ is the gyromagnetic coupling vector associated with the Magnus force, where
$Q=\frac{1}{4\pi}\int\boldsymbol{m}\cdot(\frac{\partial\boldsymbol{m}}{\partial x}\times\frac{\partial\boldsymbol{m}}{\partial y})dxdy$
is the skyrmion number~\cite{Nagaosa_NNANO2013,Zhang_JPCM2020,Gobel_PHYSREP2021}.
$\boldsymbol{\D}$ is the dissipative tensor with zero off-diagonal entries and the diagonal entries being $\D$.
$\boldsymbol{\B}$ is a term that quantifies the efficiency of the driving force.
From Eq.~(\ref{eq:TME-CPP}) we find the $\theta$-dependent $\theta_{\text{SkHE}}$~\cite{Wang_PRB2019,Xichao_PRB2022B},
\begin{equation}
\theta_{\text{SkHE}}=\arctan\left(\frac{\alpha\D\cos\theta+Q\sin\theta}{-\alpha\D\sin\theta+Q\cos\theta}\right).
\label{eq:TME-CPP-SkHE-Theta}
\end{equation}
We further computationally find that a relaxed compact skyrmion with $Q=-1$ and a diameter of about $15$ nm [Fig.~\ref{FIG1}(c)] shows $\theta_{\text{SkHE}}=158.97^{\circ}$ and no deformation during its motion driven by a small current $j=0.5$ MA cm$^{-2}$ at $\theta=0^{\circ}$, justifying its good rigidity and particle-like feature.
Hence, we obtain $\D=1.28$ and thus, find analytically from Eq.~(\ref{eq:TME-CPP-SkHE-Theta}) that $\theta_{\text{SkHE}}=0^{\circ}$ at $\theta=201.03304^{\circ}$, as shown in Fig.~\ref{FIG2}.
In the following, we set $\theta=201.03304^{\circ}$ in all computational simulations to ensure $\theta_{\text{SkHE}}=0^{\circ}$.
The skyrmions should flow in the $+x$ direction provided that they are not deformed significantly.

\subsection{The dynamic phase diagram}
\label{se:Diagram}

To explore the possible dynamic behaviors of the skyrmion pipe flow in our studied system, we first drive the skyrmions in the pipe channel into motion under a systematic variation of the current density.
We consider a uniform and a non-uniform current density distribution [Fig.~\ref{FIG1}(d)].
For the uniform case, the current density $j$ is uniform in the pipe.
For the non-uniform case [Fig.~\ref{FIG1}(e)], $j$ equals zero at the pipe edges and linearly increases to its maximum value $j_{\text{m}}$ at the middle of the lateral direction (i.e., the $y$ direction); $j$ remains constant in the longitudinal direction (i.e., the $x$ direction).
Both the uniform and non-uniform currents can drive all skyrmions into motion toward the $+x$ direction with an intrinsic skyrmion Hall angle of $\theta_{\text{SkHE}}=0^{\circ}$. Namely, we ensure that the skyrmion shows no skyrmion Hall effect (i.e., $\theta_{\text{SkHE}}=0^{\circ}$) and deformation driven by a small current.

The initial state includes $180$ relaxed compact skyrmions forming a stable triangular lattice in the pipe (Fig.~\ref{FIG1}). The $\boldsymbol{q}$ vectors of the triangular lattice are pointing at $3$, $7$, and $11$ o'clock directions.
The skyrmions fully fill the pipe and slightly interact with the upper and lower pipe edges, which are expected to form a pipe flow upon their motion toward the longitudinal direction.
Indeed, as shown in Fig.~\ref{FIG3}(a), we find that a laminar pipe flow of skyrmions could be formed when the system is driven by a small uniform or non-uniform current ($j=j_{\text{m}}\leq 10$ MA cm$^{-2}$), which will be discussed in Sec.~\ref{se:Laminar} (see \blue{Supplementary Video~1} and \blue{Supplementary Video~2}~\cite{SM}).
When the applied driving current density $j=j_{\text{m}}=10-50$ MA cm$^{-2}$, the system shows a transitional dynamic phase accompanied with transiently disordered dynamic behaviors, which will be discussed in Sec.~\ref{se:Transition} and Sec.~\ref{se:Transitional} (see \blue{Supplementary Video~3} and \blue{Supplementary Video~4}~\cite{SM}).
We also note that the skyrmions in the pipe channel will be deformed and even destroyed when the driving current is larger than a threshold value of $60$ MA cm$^{-2}$ (i.e., $j=j_{\text{m}}\geq 60$ MA cm$^{-2}$), which may lead to complex magnetic domain patterns in the pipe channel (see \blue{Supplementary Video~5} and \blue{Supplementary Video~6}~\cite{SM}).
The destruction and annihilation of skyrmions in the pipe channel driven by a uniform or non-uniform current of $j=j_{\text{m}}\geq 60$ MA cm$^{-2}$ can be seen from the time-dependent total skyrmion number $Q$ of the system [Figs.~\ref{FIG3}(b)~and~\ref{FIG3}(c)], where $Q$ approaches zero soon upon the application of the current.
In Figs.~\ref{FIG3}(d) and~\ref{FIG3}(e), the total skyrmion number of the system averaged for $500$ ns of the simulation indicates a sharp change around the dynamic phase boundary between the transitional and deformed phases. Within the deformed phase, a larger applied current density will result in the destruction and annihilation of more skyrmions.
In the following, we focus on the ordered and transiently disordered dynamic behaviors of the skyrmion flow in the pipe channel.

\subsection{Laminar dynamics and structural transition}
\label{se:Laminar}

The laminar flow means that the fluid particles flow orderly in smooth and parallel paths without lateral mixing, where each particle only has a constant velocity along the path~\cite{White_2011}. The laminar flow can be found when the fluid particles are flowing through a closed channel (e.g., a pipe).
Here we show the formation of a laminar pipe flow of skyrmions driven by a small current in the pipe.
As shown in Fig.~\ref{FIG4}, we focus on the system driven by $j=j_{\text{m}}=3$ MA cm$^{-2}$ as an example.

For the system driven by a uniform current [Figs.~\ref{FIG4}(a)-\ref{FIG4}(e)], the skyrmions move toward the $+x$ direction and the skyrmion lattice structure gradually changes to a more stable triangular configuration [Fig.~\ref{FIG4}(b)] with the three $\boldsymbol{q}$ vectors pointing at $12$, $4$, and $8$ o'clock directions (see \blue{Supplementary Video~1}~\cite{SM}).
The lattice structure transition is a result of the fact that the moving skyrmions favor a compact alignment guided by the pipe edges due to the skyrmion-skyrmion and skyrmion-edge repulsions~\cite{Xichao_PRB2022A,Xichao_PRB2022B,Souza_2023}.
When the lattice structure transition is completed, the skyrmions flow toward the $+x$ direction following nine parallel pathlines [Fig.~\ref{FIG4}(c)], where all skyrmions show the same speed, forming a moving lattice of skyrmions that can be treated as a uniform skyrmion pipe flow with a constant velocity profile. 
Such a skyrmion pipe flow does not have a free surface as the skyrmions touch and interact with both the upper and lower pipe edges.
Upon the application of the uniform current, the out-of-plane magnetization of the system $m_z$ slightly decreases, while the total skyrmion number doesn't change [Fig.~\ref{FIG4}(d)].
The decrease of $m_z$ is caused by the nature of the applied spin current with an in-plane spin-polarization direction, which tends to reduce the out-of-plane magnetization and increase the in-plane magnetization.
Therefore, as shown in Fig.~\ref{FIG5}, the average skyrmion diameter in the pipe slightly decreases upon the application of the driving current, because the spin torque with an in-plane spin-polarization direction favors more in-plane magnetization and thus leads to the increase of the circular domain wall width as well as the shrink of the out-of-plane skyrmion core.
However, a tiny deformation of the skyrmion basically does not affect $\theta_{\text{SkHE}}$, as can be seen from the skyrmion pathlines parallel to the pipe edge. The total energy suddenly increases when the current is applied and then decreases during the lattice transition; it finally approaches a stable value when a steady skyrmion pipe flow is formed.
We also note that the skyrmion size is slightly oscillating during its motion, as indicated by the time-dependent $m_z$ [Fig.~\ref{FIG4}(d)] and average skyrmion diameter (Fig.~\ref{FIG5}).
The oscillation of the skyrmion size is possibly caused by the unsteady skyrmion-skyrmion and skyrmion-edge interactions during the motion of the skyrmions driven by the current.

For the system driven by a non-uniform current [Figs.~\ref{FIG4}(f)-\ref{FIG4}(j)], the skyrmions move toward the $+x$ direction and form a laminar pipe flow (see \blue{Supplementary Video~2}~\cite{SM}).
The skyrmions flow along nine parallel pathlines [Fig.~\ref{FIG4}(h)] but with different speeds between adjacent layers of skyrmions as the skyrmion velocity $\boldsymbol{v}=(v_x,v_y)$ is proportional to $j$.
To be specific, the analytical skyrmion velocity can be obtained by Eq.~(\ref{eq:TME-CPP}) as
$v_{x}=\left|u\right|\I\frac{Q\cos\theta-\alpha\D\sin\theta}{Q^2+\alpha^2\D^2}$
and $v_y=\left|u\right|\I\frac{\alpha\D\cos\theta+Q\sin\theta}{Q^2+\alpha^2\D^2}$, where
$\I=\frac{1}{4\pi}\iint(\frac{\partial\boldsymbol{m}}{\partial x}\times\boldsymbol{m})_{y}dxdy$
is related to the efficiency of the spin torque over the skyrmion~\cite{Xichao_PRB2022B,Wang_PRB2019}.
When $\theta_{\text{SkHE}}=0^{\circ}$ ($\theta=201.03304^{\circ}$), the skyrmion velocity $v_{x}\sim cj$ and $v_y\sim 0$ with $c$ being a coefficient as long as $u\sim j$ and the skyrmion is not obviously deformed.
In the fully developed laminar flow of skyrmions, the skyrmions flow orderly following parallel pathlines without obvious lateral motion, and therefore, they show a dynamically varying lattice structure, where one may find mixed square and triangular lattice structures at selected times [Fig.~\ref{FIG4}(g)].
Due to the dynamically varying lattice structure, the skyrmion-skyrmion interactions lead to more pronounced oscillations in $m_z$ and total energy of the system, however, the total skyrmion number remains constant [Figs.~\ref{FIG4}(i)~and~\ref{FIG4}(j)].
As shown in Fig.~\ref{FIG5}, the oscillation of the skyrmion size driven by the non-uniform current is also more pronounced than that driven by the uniform current due to the varying spacing between neighboring skyrmions in the dynamically varying skyrmion lattice.

\subsection{Transition from pipe flow to open-channel flow}
\label{se:Transition}

In fluid dynamics, a laminar flow may transform into a transitional flow and further to a turbulent flow as the Reynolds number increases. The speed of the fluid plays an important role on the transition as the Reynolds number increases with the flow speed.
To explore similar phenomena in the skyrmion pipe flow, we apply a moderately large current to drive the system as the skyrmion speed increases with $j$.
For the system driven by a uniform current, we do not find the laminar-turbulent transition for a moderate range of $j=10-50$ MA cm$^{-2}$, however, we observe a transition of the skyrmion pipe flow into an open-channel flow due to the skyrmion deformation-induced skyrmion Hall effect (see \blue{Supplementary Video~3}~\cite{SM}).
In Fig.~\ref{FIG6}, we show the system driven by $j=40$ MA cm$^{-2}$ as an example.

In principle, the skyrmions should move with $\theta_{\text{SkHE}}=0^{\circ}$ as shown in Fig.~\ref{FIG4}.
However, when a moderately large current is applied, the skyrmion shows certain but not significant deformation and its size is slightly increased~\cite{Litzius_NPHYS2017}, indicated by the sharp decrease of $m_z$ in Fig.~\ref{FIG7}(a) and the sharp increase of the average skyrmion diameter in Fig.~\ref{FIG8}.
As $\theta_{\text{SkHE}}\sim\D$ [Eq.~(\ref{eq:TME-CPP-SkHE-Theta})], while $\D=\pi{^2}d/8\gamma_{\text{DW}}$~\cite{Wanjun_NPHYS2017} with $d$ and $\gamma_{\text{DW}}$ being the skyrmion diameter and the domain wall width, respectively, the increase of the skyrmion size could lead to the lateral motion of the skyrmions (i.e., a non-zero $\theta_{\text{SkHE}}$) [Fig.~\ref{FIG7}(c)].
The motion of the skyrmions toward the lower pipe edge results in the compression of the skyrmions~\cite{Reichhardt_PRB2020,Souza_2023}, which leads to the formation of a skyrmion flow with a free surface and reduced flow width.
Due to the compression effect, the skyrmion size decreases as indicated by the gradual increase of $m_z$ in Fig.~\ref{FIG7}(a) as well as the gradual decrease of the average skyrmion diameter in Fig.~\ref{FIG8}.
Hence, $\theta_{\text{SkHE}}$ equals zero when the skyrmion size almost decreases to its initial value.
The compressed skyrmions flow toward the $+x$ direction and follow eight parallel pathlines [Fig.~\ref{FIG7}(d)].
No skyrmion is annihilated during the compression [Fig.~\ref{FIG7}(a)], and the system reaches a higher total energy when a steady open-channel flow of skyrmions is formed [Fig.~\ref{FIG7}(b)].
The speed of the skyrmions on the free surface is faster than that of the main skyrmion flow, which forms a single shear layer of skyrmions at the flow surface~\cite{Souza_2023,Reichhardt_PRB2020}.
It should noted that the effect of the moderately large driving current in narrowing the width of the skyrmion flow depends on the applied current density, as shown in Fig.~\ref{FIG3}(a). Namely, a larger current density could result in a narrower usable portion of the pipe channel; however, it will also affect the density and crystal structure of the skyrmion flow.

\subsection{Transitional dynamics}
\label{se:Transitional}

For the system driven by a moderately large non-uniform current, we do not find a fully developed turbulent flow, however, we find both laminar and transiently disordered dynamic behaviors of skyrmions in a transitional skyrmion flow (see \blue{Supplementary Video~4}~\cite{SM}).
In fluid dynamics, a transitional flow is a mixed flow with both laminar and turbulent dynamics, usually with turbulent flow in the pipe center, and laminar flow near the pipe edges.
In contrast, taking the system driven by $j_{\text{m}}=40$ MA cm$^{-2}$ as an example (Figs.~\ref{FIG9}), we find that the skyrmions close to the lower pipe edge show laminar dynamics, while those in the upper half of the pipe could show disordered behaviors.
The reason is that the relatively large $j$ in the pipe center (i.e., near $y=250$ nm) results in the deformation of skyrmions, which further leads to non-zero $\theta_{\text{SkHE}}$ and compression effect similar to the situation given in Fig.~\ref{FIG6}.
The skyrmion traveling toward the $+x$ direction with non-zero $\theta_{\text{SkHE}}$ shows a lateral motion toward the $+y$ or $-y$ direction, depending on its deformation (e.g., current-induced expansion or compression-induced shrink), as indicated by the time-dependent variations of $m_z$ [Fig.~\ref{FIG10}(a)], the total energy [Fig.~\ref{FIG10}(b)], and the average skyrmion diameter (Fig.~\ref{FIG11}).
The transiently disordered motion of skyrmions is prominent in the upper half of the skyrmion flow, demonstrated by the irregular pathlines [Figs.~\ref{FIG10}(c)~and~\ref{FIG10}(d)].
However, the skyrmions adjacent to the upper and lower pipe edges show stable laminar dynamics driven by relatively small $j$.
Interestingly, a single skyrmion is annihilated in the transitional flow of skyrmions as indicated by the change of $Q$ in Fig.~\ref{FIG10}(a) near $t=310$ ns, which may be collapsed due to the strong compressive interactions between skyrmions in the transiently disordered region.
We note that the strong compressive interactions between skyrmions that lead to the shrink and collapse of a skyrmion are also indicated by an obvious peak of the average skyrmion diameter near $t=310$ ns.
After the annihilation of the skyrmion, the transitional flow gradually transforms into a laminar flow, where $10$ layers of skyrmions flow following parallel pathlines toward the $+x$ direction [Fig.~\ref{FIG10}(e)].

Here, it is worth mentioning that the oscillation of the skyrmion size driven by a moderately large current (Figs.~\ref{FIG8}~and~\ref{FIG11}) is not obvious compared to that driven by a small current (Fig.~\ref{FIG5}), because a larger spin current has a stronger effect on the profile of a skyrmion, which could suppress the skyrmion oscillation and even lead to the skyrmion deformation.

\subsection{Effect of magnetic anisotropy variation}
\label{se:Anisotropy}

In this section, we further investigate the laminar dynamics of the skyrmion pipe flow in the presence of certain magnetic anisotropy variation.
The model geometry and parameters are the same as that used in Sec.~\ref{se:Laminar}, however, a variation of the magnetic anisotropy constant in the pipe channel is applied by considering grain-like regions with slightly varying PMA constants using the Voronoi tessellation~\cite{MuMax}.
To be specific, we define grains with the Voronoi tessellation over the pipe channel with a grain size of $100$ nm, a maximum region number of $50$, and a random seed of $777$. We consider random $0.5\%-2.0\%$ anisotropy constant variation within all grain regions with the default value being $K=0.8$ MJ m$^{-3}$.

As shown in Fig.~\ref{FIG12}, we first study the system driven by a small uniform current of $j=3$ MA cm$^{-2}$. It shows that the variation of magnetic anisotropy in the pipe channel introduces certain pinning and defect effects, which lead to irregular distribution of skyrmions moving in the pipe.
The skyrmions show obvious longitudinal and transverse motion driven by the uniform current even at a $0.5\%$ variation of the magnetic anisotropy (see \blue{Supplementary Video~7}~\cite{SM}), which is different from the case in the clean pipe channel [Fig.~\ref{FIG4}(a)] where all skyrmions move toward the longitudinal direction (i.e., the $+x$ direction).
When a higher $2.0\%$ variation of the magnetic anisotropy is applied in the pipe channel, the skyrmions show more irregular motion (see \blue{Supplementary Video~8}~\cite{SM}). Consequently, the effect of the magnetic anisotropy variation leads to multiple wavy pathlines of the skyrmions in the pipe, as shown in Figs.~\ref{FIG13}(a)-\ref{FIG13}(d).
We note that some skyrmions are annihilated due to the strong compression of skyrmions at grain boundaries when a $2.0\%$ variation of the magnetic anisotropy is applied, as indicated by the time-dependent $Q$ given in Fig.~\ref{FIG13}(e).

In Fig.~\ref{FIG13}(f), we show the time-dependent $m_z$ for the systems with different variations of the magnetic anisotropy. It can be seen that $m_z$ increases with time and approaches a certain value especially for the system with a higher variation of the magnetic anisotropy. This is in contrast to the case without the variation of the magnetic anisotropy [cf. Fig.~\ref{FIG4}(d) and Fig.~\ref{FIG13}(f)], where $m_z$ slightly decreases with time and approaches a certain value due to the in-plane polarized spin current. The reason is that the skyrmion is difficult to penetrate into a grain with a relatively higher anisotropy~\cite{Ohara_NL2021}. Therefore, in the pipe channel with higher variation of the magnetic anisotropy, the skyrmions will experience stronger compression effect near the potential barriers formed by the boundaries of grains with enhanced anisotropy. As a result, the current-induced compression of skyrmions near the grain boundaries leads to the decrease of the skyrmion size (Fig.~\ref{FIG14}) as well as the annihilation of the skyrmions [Fig.~\ref{FIG13}(e)]. Both the decrease of the skyrmion size and the decrease of the number of skyrmions in the pipe channel result in the increase of $m_z$.

It should be noted that the effect of the anisotropy variation on the skyrmion pipe flow driven by a small non-uniform current (Figs.~\ref{FIG15}-\ref{FIG17}) is qualitatively similar to that driven by a small uniform current (Figs.~\ref{FIG12}-\ref{FIG14}).
Namely, a higher variation could generate more complicated and irregular pathlines of skyrmions moving in the pipe (Figs.~\ref{FIG15}~and~\ref{FIG16}), and may also result in the annihilation of certain skyrmions [Fig.~\ref{FIG16}(e)].
For example, see \blue{Supplementary Video~9} and \blue{Supplementary Video~10}~\cite{SM} for the dynamic behaviors of skyrmions driven by a small non-uniform current ($j_{\text{m}}=3$ MA cm$^{-2}$) in the pipe channel with a $0.5\%$ and $2.0\%$ anisotropy variations, respectively.

\section{Conclusion}
\label{se:Conclusion}

In conclusion, we have studied the flow dynamics of skyrmions in a 2D pipe driven by small and moderately large currents.
A lattice structural transition may happen due to the skyrmion-edge and skyrmion-skyrmion interactions.
A small non-uniform current could create a laminar skyrmion flow, while a large non-uniform current may create a transitional skyrmion flow.
A uniform current may also lead to a transition of the skyrmion pipe flow to an open-channel flow.
Namely, the width of the skyrmion flow could be controlled by the current density of the applied uniform current.
A chain of skyrmions on the free surface of the open-channel skyrmion flow could move faster than the main skyrmion flow.
Our results reveal the rich dynamics of fluid-like topological spin textures.
Our results may also motivate future research toward the understanding of the complex flow and transport phenomena in magnets.


\begin{acknowledgments}
X.Z. and M.M. acknowledge support by CREST, the Japan Science and Technology Agency (Grant No. JPMJCR20T1).
M.M. also acknowledges support by the Grants-in-Aid for Scientific Research from JSPS KAKENHI (Grant No. JP20H00337).
J.X. was a JSPS International Research Fellow supported by JSPS KAKENHI (Grant No. JP22F22061).
O.A.T. acknowledges support by the Australian Research Council (Grant No. DP200101027), the Cooperative Research Project Program at the Research Institute of Electrical Communication, Tohoku University (Japan), and by the NCMAS grant.
M.E. acknowledges support by CREST, JST (Grant No. JPMJCR20T2).
G.Z. acknowledges support by the National Natural Science Foundation of China (Grants No. 51771127, No. 51571126, and No. 51772004), and Central Government Funds of Guiding Local Scientific and Technological Development for Sichuan Province (Grant No. 2021ZYD0025).
Y.Z. acknowledges support by the National Natural Science Foundation of China (Grants No. 11974298 and No. 12374123), the Shenzhen Fundamental Research Fund (Grant No. JCYJ20210324120213037), the Shenzhen Peacock Group Plan (Grant No. KQTD20180413181702403), and the Guangdong Basic and Applied Basic Research Foundation (Grant No. 2021B1515120047).
X.L. acknowledges support by the Grants-in-Aid for Scientific Research from JSPS KAKENHI (Grants No. JP20F20363, No. JP21H01364, No. JP21K18872, and No. JP22F22061).
\end{acknowledgments}



\end{document}